\documentclass[aps,prx,reprint,onecolumn,12pt, preprintnumbers,longbibliography,superscriptaddress,notitlepage,floatfix]{revtex4-2}
\usepackage{color,soul}
\usepackage{graphicx}
\usepackage{dcolumn}
\usepackage{bm}
\usepackage[english]{babel}
\usepackage{hyperref}
\hypersetup{pdfstartview={FitH},pdfpagemode={UseNone},
            colorlinks,linkcolor=blue, citecolor=blue, urlcolor=blue,
            bookmarksopen=true, pdfnewwindow=true}
\usepackage[all]{hypcap}

\usepackage{siunitx}
\renewcommand{\thesection}{\arabic{section}}
\renewcommand{\thesubsection}{\arabic{section}.\arabic{subsection}}

\begin{document}
\title{Spatially entangled photon-pairs from lithium niobate nonlocal metasurfaces}

\author{Jihua Zhang}
\altaffiliation{Co-first authors with equal contribution}

\author{Jinyong Ma}
\altaffiliation{Co-first authors with equal contribution}

\author{Matthew Parry}
\author{Marcus Cai} 
\author{Rocio~Camacho~Morales}
\affiliation{ARC Centre of Excellence for Transformative Meta-Optical Systems (TMOS), Research School of Physics, The~Australian National University, Canberra, ACT 2601, Australia}

\author{Lei Xu}
\affiliation{Advanced Optics and Photonics Laboratory, Department of Engineering, School of Science and Technology, Nottingham Trent University, Nottingham, NG11 8NS, UK}
\author{Dragomir N. Neshev}
\author{Andrey A. Sukhorukov} 
\email{andrey.sukhorukov@anu.edu.au}
\affiliation{ARC Centre of Excellence for Transformative Meta-Optical Systems (TMOS), Research School of Physics, The~Australian National University, Canberra, ACT 2601, Australia}


\begin{abstract}
\noindent\textbf{ABSTRACT} 
Metasurfaces consisting of nano-scale structures are underpinning new physical principles for the creation and shaping of quantum states of light. Multi-photon states that are entangled in spatial or angular domains are an essential resource for quantum imaging and sensing applications, however their production traditionally relies on bulky nonlinear crystals. We predict and demonstrate experimentally the generation of spatially entangled photon pairs through spontaneous parametric down-conversion from a metasurface incorporating a nonlinear thin film of lithium niobate. This is achieved through nonlocal resonances with tailored angular dispersion mediated by an integrated silica meta-grating, enabling control of the emission pattern and associated quantum states of photon pairs by designing the grating profile and tuning the pump frequency. We measure the correlations of photon positions and identify their spatial anti-bunching through violation of the classical Cauchy-Schwartz inequality, witnessing the presence of multi-mode entanglement. Simultaneously, the photon-pair rate is strongly enhanced by 450 times as compared to unpatterned films due to high-quality-factor metasurface resonances, and the coincidence to accidental ratio reaches 5000. These results pave the way to miniaturization of various quantum devices by incorporating ultra-thin metasurfaces functioning as room-temperature sources of quantum-entangled photons.

\end{abstract}


\maketitle

\section*{Introduction}
Quantum entanglement underpins a broad range of fundamental physical effects~\cite{Horodecki:2009-865:RMP} and serves as an essential resource in various applications
including quantum imaging~\cite{Shih:2007-1016:ISQE},  communications~\cite{Gisin:2007-165:NPHOT}, information processing and computations~\cite{Bennett:2000-247:NAT}. In optics, the most common source of entangled photons is based on the spontaneous parametric down-conversion (SPDC) process in quadratically nonlinear materials~\cite{Klyshko:1988:PhotonsNonlinear}, which can operate at room temperature. The generated photons can be entangled in
transverse~\cite{Chan:2007-50101:PRA, Just:2013-83015:NJP} and orbital angular momenta~\cite{Mair:2001-313:NAT}, effectively accessing a large Hilbert space~\cite{Law:2004-127903:PRL, Torres:2003-50301:PRA, Salakhutdinov:2012-173604:PRL}. 
Strong transverse momentum entanglement was recently realized with a micrometer-scale film of nonlinear material lithium niobate (LiNbO$_3$)~\cite{Okoth:2020-11801:PRA}, yet the compactness came at a cost of a strongly reduced generation rate. This leads to a fundamentally and practically important research question on the potential for efficient generation of spatially entangled photons in ultra-thin optical structures.

Over the last decade, dramatic enhancements of nonlinear light-matter interactions were achieved in nanofabricated structures with subwavelength thickness, known as metasurfaces~\cite{Huang:2020-126101:RPP, DeAngelis:2020:NonlinearMetaOptics}, which are also bringing advances
to the field of quantum optics~\cite{Solntsev:2021-327:NPHOT}. 
An enhancement of SPDC through localized Mie-type optical resonances~\cite{Kuznetsov:2016-aag2472:SCI} in nanoantennas~\cite{Marino:2019-1416:OPT} and metasurfaces~\cite{Santiago-Cruz:2021-4423:NANL} was experimentally demonstrated. 
Recent theoretical studies~\cite{Parry:2021-55001:ADP, Jin:2021-19903:NASC, Mazzanti:2022-35006:NJP} suggested that using metasurfaces with nonlocal lattice resonances can further boost the SPDC efficiency, yet the realization of this concept remained outstanding.

In this work, we demonstrate experimentally, for the first time to our knowledge, that strongly enhanced generation of spatially entangled photon pairs can be achieved from metasurfaces supporting nonlocal 
resonances at the signal and idler in the telecommunication band around the 1570~nm wavelength. The SPDC enhancement of $\sim$ 450 times compared with unpatterned film and the coincidence-to-accidental ratio (CAR) of $\sim5000$ are an order of magnitude higher than what has been possible to date~\cite{Santiago-Cruz:2021-4423:NANL}, benefiting from the nonlocal feature of the resonances. This is the foundation for the preparation of strongly entangled quantum states with a much higher spectral and spatial brightness compared to localized resonances.
Most importantly, our experiments indicate spatial entanglement of photon pairs by violating the classical Cauchy-Schwartz inequality (CSI), 
confirming the practical path for the preparation of high-quality entanglement sources with metasurfaces.

\section*{Results}
\subsection*{Concept and modelling} 

\begin{figure*}[!b] \centering
\includegraphics[width=1\textwidth]{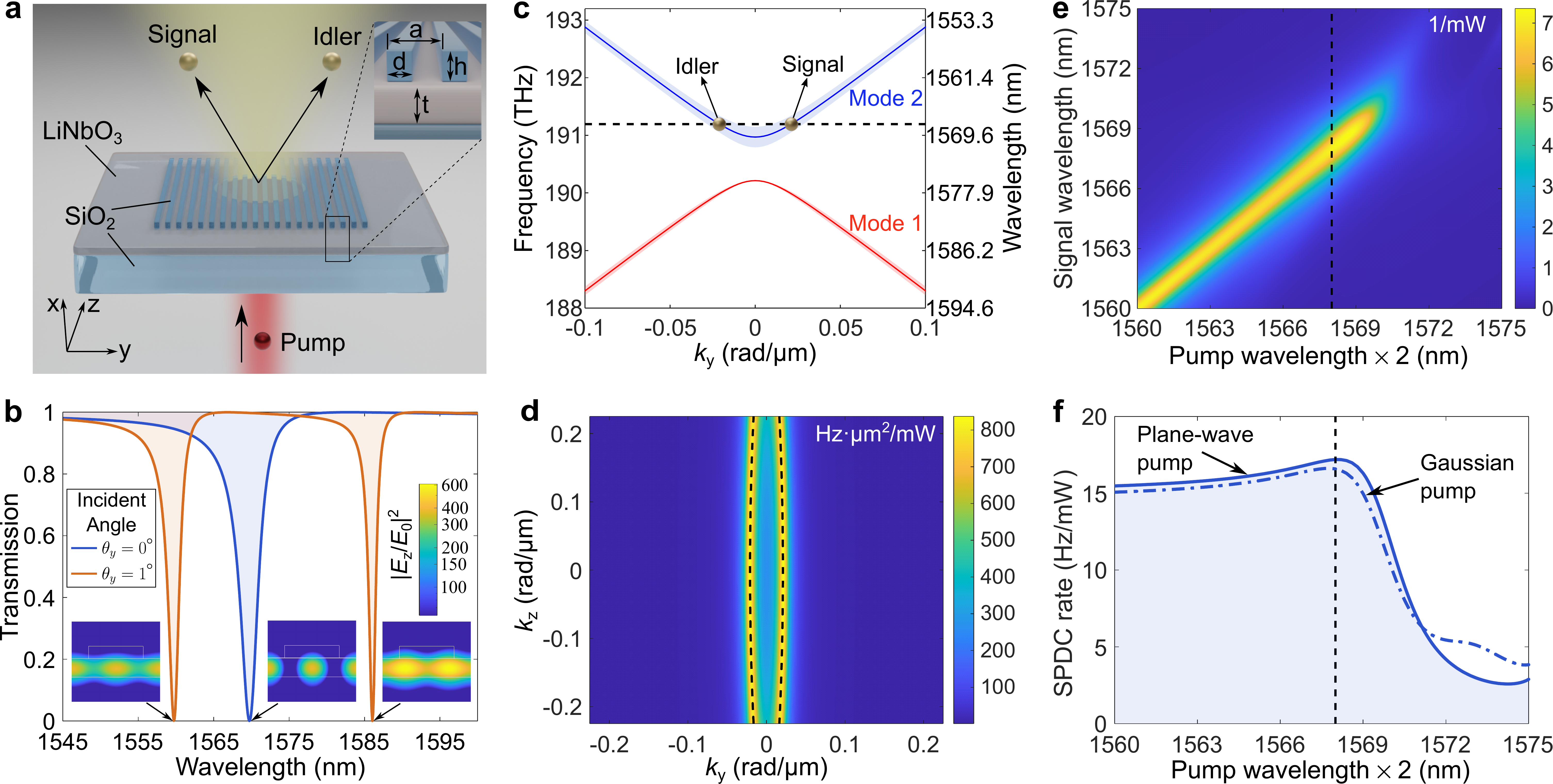} 
\caption{\textbf{Nonlocal metasurfaces for efficient generation of spatially entangled photon pairs.} \textbf{a} Sketch of spatially entangled signal and idler photons generation from a LiNbO$_3$ thin film covered by a $\text{SiO}_2$ grating and pumped by a continuous laser. The optical axis of LiNbO$_3$ and the grating are along the $z$ direction. Inset shows the dimensions of the metasurface (not to scale), which we choose as $a=890$~nm, $d=550$~nm, $h=200$~nm, and $t=304$~nm. \textbf{b} Simulated transmission spectra of the metasurface for $z$-polarized light, showing a single guided-mode resonance at normal incidence and two resonances with a non-zero incident angle. This angular dependent transmission features a nonlocal response. The insets are the field intensity at the resonances, showing strong field enhancement in the LiNbO$_3$ layer. \textbf{c}~Simulated eigenfrequencies of resonances vs. the transverse wavenumber $k_y$ at $k_z=0$. The shaded regions represent the bandwidth of the resonances. Signal \& idler photons produced symmetrically relative to the $\Gamma$ point satisfy both the energy and transverse momentum conservation, when half the pump frequency $\omega_{\rm p}$ is within the resonance range. The black dashed line marks $\omega_{\rm p}/2=191.19$~THz, which is slightly blue detuned from the normal incidence resonance of the bright mode~2. \textbf{d} Predicted photon emission rate at the marked pump frequency as a function of transverse wavenumbers. The black dashed line indicates the momentum matching conditions for the degenerate SPDC. \textbf{e} Spectra of the signal photons for different pump frequencies. \textbf{f} The total emission rate after integrating over the signal spectrum for plane-wave and Gaussian ($100 \mu m$ diameter) pump beam. The black dashed lines in (e,f) relate to the pump frequency marked in~\textbf{c}.}
\label{fig:theory}
\end{figure*}


We develop a nonlocal metasurface based on the 
LiNbO$_3$-on-insulator
platform, which was recently employed for various applications including electro-optic modulation~\cite{Klopfer:2020-5127:NANL, Weiss:2022-605:ACSP} and classical optical frequency conversion~\cite{Duong:2109.08489:ARXIV}.
The LiNbO$_3$ material features large second-order susceptibility, low fluorescence, and high optical transmission in a broad wavelength range~\cite{Saravi:2021-2100789:ADOM, Zhu:2021-242:ADOP}, which is essential for the quality of generated photon pairs.
We design
a periodic $\text{SiO}_2$ grating on top of a lithium niobate film with sub-wavelength thickness ($t \simeq 304$~nm), as schematically illustrated in Fig.~\ref{fig:theory}\textbf{a}. 
%
We select the $x$-cut of LiNbO$_3$ and orient the grating along the \textit{z} axis of the film, such that 
the efficiency of SPDC 
is maximized when the pump beam and emitted photons are linearly polarized along the \textit{z} axis, because of its highest nonlinear tensor coefficient $\chi^{(2)}_{\rm zzz}$. 
Note that this design does not require nanopatterning of the LiNbO$_3$ film in contrast to Mie-resonant metasurfaces~\cite{Santiago-Cruz:2021-4423:NANL}, thereby avoiding possible damage at the edges while preserving the total volume of the nonlinear material.

The grating supports guided-mode resonances inside the LiNbO$_3$ layer when the transverse wavenumber introduced by the incident wave and the grating matches the propagation wavenumber of the slab modes. This physical mechanism underpins a strongly nonlocal response, since the resonant excitations can spread out in-plane through the guided waves~\cite{Zhang:2020-76:LSA, Song:2021-1224:NNANO, Kwon:2018-173004:PRL}. Mathematically, the nonlocality in space is intrinsically linked to the presence of angular dispersion. 
We first discuss the key features by considering a case of the incident plane wave which transverse wavenumber along the grating is vanishing, i.e. $k_z=0$. In the weak-grating regime, which does not exactly apply to metasurfaces yet captures the essential physics, the first-order resonance happens when the following condition is satisfied \cite{Wang:1993-2606:AOP, Quaranta:2018-1800017:LPR}:
\begin{equation} \label{eq:grating}
    \frac{2\pi}{a} \pm~\sin\theta_y \frac{2\pi}{\lambda} = n_{\rm eff} \frac{2\pi}{\lambda} \,,
\end{equation}
%
where $a$ is the period of the grating, $\theta_y$ is the incident angle of the input plane wave in the $y-x$ plane, $\lambda$ is the wavelength in vacuum, $n_{\rm eff}$ is the effective index of the waveguide mode in the LiNbO$_3$ film, and the signs $\pm$ correspond to the two guided modes propagating in opposite directions along the $y$ axis. Equation~(\ref{eq:grating}) indicates the presence of two resonant wavelengths at $\lambda=a(n_{\rm eff}\pm~\sin\theta_y)$, which depend on the incident 
angle, and can be 
controlled by selecting the period of the grating.

We determine an approximate grating period for the target photon-pair wavelengths in the telecommunication band using Eq.~(\ref{eq:grating}), and then fine-tune the structure parameters by performing finite-element modelling of Maxwell's equations to also facilitate electromagnetic field localization inside the LiNbO$_3$ layer. 
The optimized geometry is sketched in the inset of Fig.~\ref{fig:theory}\textbf{a}. 
We present the characteristic transmission spectra in Fig.~\ref{fig:theory}\textbf{b} at normal and tilted incidence as indicated by labels, and show the dependencies of the resonant eigenfrequencies ($\omega$) and bandwidths defined by the loss coefficients ($2\gamma$) of the quasi-normal modes on the transverse wavenumbers along the $y$-axis in Fig.~\ref{fig:theory}\textbf{c}. The corresponding quality factors of the resonances are $Q = \omega / (2 \gamma)$.

Let us first analyse the mode features at the $\Gamma$ point, i.e., $k_y = 0$. There is a small frequency splitting due to a second-order grating scattering that is not captured by simplified Eq.~(\ref{eq:grating}).
The lower-frequency mode 1 has zero bandwidth and thus an infinite quality factor. We check that this mode has an anti-symmetric electric field profile (see Supplementary Fig.~S1), indicating its origin as a symmetry-protected extended bound state in the continuum (BIC)~\cite{Hsu:2016-16048:NRM}. 
The single resonant transmission dip at normal incidence (Fig.~\ref{fig:theory}\textbf{b}) corresponds to an excitation of a bright mode 2, which has a quality factor of $\simeq 500$.
As shown in the inset, the mode has a symmetric standing-wave intensity profile due to the interference of equally excited counter-propagating guided modes. 



For an incident angle of $1^\circ$, with $k_y \simeq 0.07\,{\rm rad}/\mu {\rm m}$, two transmission dips appear at both sides of the resonance at $k_y=0$. The weakly modulated intensity profiles shown in the insets of Fig.~\ref{fig:theory}\textbf{b} indicate the dominance of one guided mode, in agreement with the 
general properties of lattice resonances.
For both mode 1 and mode 2, strong intensity enhancement of up to 600 times is observed inside the LiNbO$_3$ layer,
which can accordingly increase the efficiency of the nonlinear processes including second harmonic generation (SHG) and SPDC as we demonstrate in the following. 

Most importantly, the high Q-factors are retained for a broad range of incidence angles, while the two resonance frequencies move further apart for the larger values of $|k_y|$.
This dispersion dependence defines the group velocity of the quasi-guided modes in the lithium niobate layer, $\nu_g = \partial \omega / \partial k_y$. Noting that the characteristic mode lifetime where the intensity decreases by a factor of two can be found as $\tau \simeq {\rm ln}(2) / (2 \gamma) = {\rm ln}(2)\,Q / \omega$,
we estimate the propagation distance in $y$-direction of the quasi-guided mode~2 at  $k_y \simeq 0.07\,{\rm rad}/\mu {\rm m}$ as $\tau \nu_g \simeq 71 \mu m$. The latter value represents approximately $80$ grating periods, confirming a strongly nonlocal nature of the metasurface resonances.

In addition to the high-quality resonances, both the energy and transverse momentum conservation conditions need to be satisfied in the SPDC process of photon-pair generation.
We designed the meta-grating to satisfy an in-plane symmetry $y \rightarrow -y$, such that the resonances at $k_y$ and $-k_y$ appear at the same frequency.
This property allows the simultaneous fulfillment of phase and energy matching of a spectral-degenerate SPDC process when a normally incident pump has the half-frequency in the guided mode resonance range. For example, the black dashed line in Fig.~\ref{fig:theory}\textbf{b} corresponds to a pump frequency $\omega_p=2\times191.19$~THz. Accordingly, the frequency of the resonantly enhanced signal and idler photons and their emission angles can be controlled by tuning the pump frequency, which allows one to tailor the spectrum and spatial entanglement of the photon pairs.


 

The SPDC emission can occur over a range of transverse momenta and photon frequencies, and accordingly the associated quantum bi-photon states belong to a high-dimensional Hilbert space. Their modelling requires fast and accurate simulation methods. For this purpose, we adopt a coupled mode theory (CMT)~\cite{Suh:2004-1511:IQE, Sun:2201.08156:ARXIV} approach to our metasurface design, and verify that it precisely describes the metasurface resonances, including all the angular and frequency features (see Supplementary Sec.~S1).
We then use the CMT to efficiently simulate the sum frequency generation (SFG) process and calculate the SPDC emission via quantum-classical correspondence~\cite{Marino:2019-1416:OPT, Parry:2021-55001:ADP}. Figure~\ref{fig:theory}\textbf{d} shows the predicted photon-pair rate integrated over their frequency spectra vs. the transverse momenta in the metasurface plane, for a normally incident plane-wave pump with the frequency 
$\omega_p=2\times191.19$~THz. 
The elongated emission pattern in the $z$-direction reflects a weak quadratic dispersion dependence on $k_z$ (see Supplementary Fig.~S2). We mark with the black dashed line the transverse phase-matching condition at the centre of mode resonances, which aligns with the emission peaks.

We present the dependence on the pump frequency of the photon spectra integrated over all emission angles in Fig.~\ref{fig:theory}\textbf{e}, and the total emission rate  in Fig.~\ref{fig:theory}\textbf{f}. Here the frequency range corresponding to the excitation of mode~2 is shown, while similar features are observed for mode~1 away from the band-edge, where tunable off-normal photon emission can be achieved.
The black dashed lines correspond to the pump frequency $\omega_p=2\times191.19$~THz, at which the highest total rate is predicted. The corresponding degenerate photon frequencies are slightly blue detuned from the bright mode~2 resonance at normal incidence, 
since in the latter case the Q-factor is lower and quadratic band-edge dispersion affects the phase-matching.
We find that the SPDC rate can be enhanced by over two orders of magnitude compared with an unpatterned LiNbO$_3$ film of the same thickness, with an even stronger increase of the spectral brightness (see Supplementary Fig.~S4).
Importantly, the enhancement is preserved under the practical experimental conditions of a Gaussian beam pump rather than an idealised plane-wave, 
as shown in Fig.~\ref{fig:theory}\textbf{f}. 
%


It is a remarkable feature that the enhanced photon rate stays practically constant as the pump wavelength is decreased. 
The tuning range of the signal and idler wavelengths, determined by twice the pump wavelength, can be up to hundreds of nanometers without temperature adjustment. 
These properties are facilitated by the nonlocal metasurface resonances, which dispersion also mediates the tunability of the emission angles, offering great flexibility for future applications.



\subsection*{Experimental characterization: metasurface resonances and enhanced second-harmonic generation} 

\begin{figure*}[!b] \centering
\vspace*{10mm}
\includegraphics[width=1\textwidth]{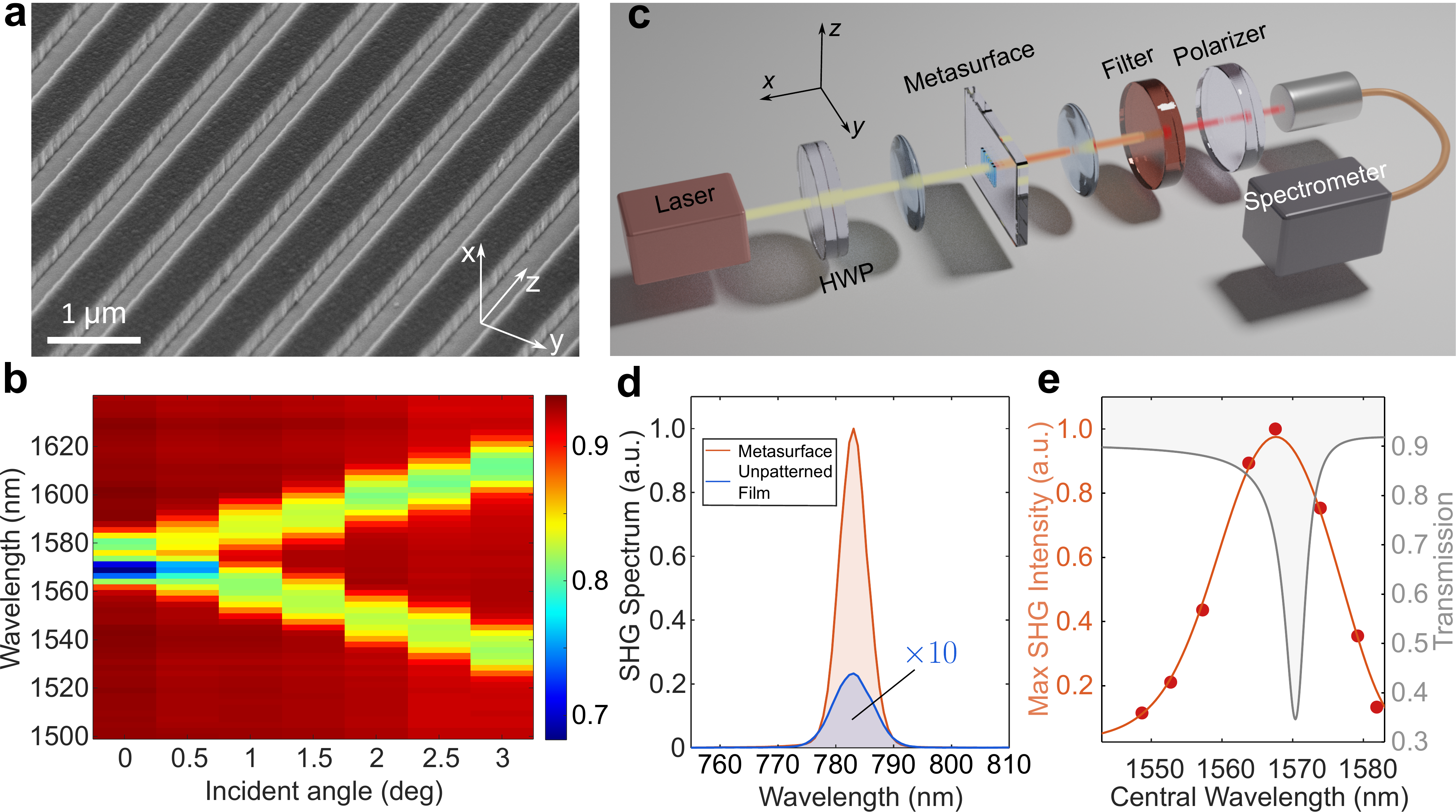} 
\caption{\textbf{Classical experiment: linear transmission and enhanced second-harmonic generation of the nonlocal metasurface.} \textbf{a} SEM image of the fabricated dielectric meta-grating on top of an \textit{x}-cut 304~nm thick lithium niobate film. \textbf{b}~Measured linear transmission as a function of incident angle and wavelength. The resonance is nearly-degenerate at normal incidence and shows a mode splitting at a larger angle. The transmission is measured with a tungsten-halogen broadband lamp. \textbf{c} Experimental setup for second-harmonic generation. The metasurface is pumped with femtosecond laser pulses from the grating side, with a tunable central wavelength and a bandwidth of 23~nm. The SHG signal is collected with a spectrometer. \textbf{d} SHG spectra from metasurface (blue curve) and unpatterned film (red curve). The SHG intensity is normalized to its maximum from the metasurface.  \textbf{e} Normalized peak SHG intensity vs. the central wavelength of the input pulses. The shaded grey curve shows the Fano fitting of the metasurface transmission measured with a tunable CW laser (1500 nm to 1575 nm) at normal incidence (See Supplementary Fig. S6). This measurement shows a narrower band but a higher resolution as compared with the results obtained from the lamp. The red dots are max SHG intensity extracted from SHG spectra obtained at different pulsed laser central wavelengths and the red line is the corresponding spline fitting.
The optimal enhancement is found near the optical resonance. 
}
\label{fig:classical}
\end{figure*}

We fabricate the metasurface by electron beam lithography and etching processes as described in the Methods section. The scanning electron microscopy (SEM) image of the meta-grating is shown in Fig.~\ref{fig:classical}(a), which confirms that the dimensions closely match the optimal design geometry according to our theoretical analysis.
The experimental transmission measurements [Fig.~\ref{fig:classical}\textbf{b}] of the metasurface identify a resonance at 1570.5~nm with a bandwidth of 3.5~nm and a quality factor $Q \sim 455$, for a normally incident light. Note that another shallow dip at 1580~nm corresponding to the dark mode~1
resonance is also visible at normal incidence, because fabrication imperfections break the symmetry and transform the mode into a quasi-BIC one. Two resonances with a visible spectral splitting manifest at a non-zero incidence angle. The observed angular dispersion confirms the nonlocal origin of resonances in agreement with the theoretical modelling presented above. 

The classical nonlinear effects, second-harmonic  (SHG) and sum-frequency generation (SFG), can be regarded as the reverse processes of the SPDC~\cite{Marino:2019-1416:OPT, Parry:2021-55001:ADP}. It was recently established that SHG and SFG can occur in LiNbO$_3$ thin films~\cite{Ma:2020-145:OL, ZhuInfrared2021} and can be enhanced in metasurfaces with Mie-type resonances~\cite{Kim:2018-4769:ACSP, Fedotova:2020-8608:NANL, Carletti:2021-731:ACSP, Ma:2021-2000521:LPR}. Here we test the nonlinearity enhancement by exploring the SHG from the nonlocal metasurface, using the experimental setup sketched in Fig.~\ref{fig:classical}\textbf{c}. The SHG is triggered with a femtosecond laser whose central wavelength is tunable from 1540 to 1580~nm. In all SHG experiments, the laser is normally incident onto the sample from the grating side.

We compare the efficiency of the SHG from the metasurface and unpatterned film in Fig.~\ref{fig:classical}\textbf{d}, and register up to 50-times enhancement.
Note that the pulsed laser bandwidth is 23~nm, which is larger than the linewidth of the metasurface resonances, significantly limiting the measured enhancement. Additionally, we characterize the SHG enhancement by laser pulses with different central wavelengths, see Fig.~\ref{fig:classical}\textbf{e}. The optimal enhancement is observed when the pulse central wavelength sits within the bright-mode resonance of the metasurface. These results confirm the high quality of the fabricated metasurfaces, which strongly boost nonlinear wave mixing in the LiNbO$_3$ layer due to nonlocal resonances according to our theoretical concept.


\subsection*{Quantum measurements: enhanced photon-pair generation and spatial entanglement}

\begin{figure*}[!b] \centering
\vspace*{5mm}
\includegraphics[width=1\textwidth]{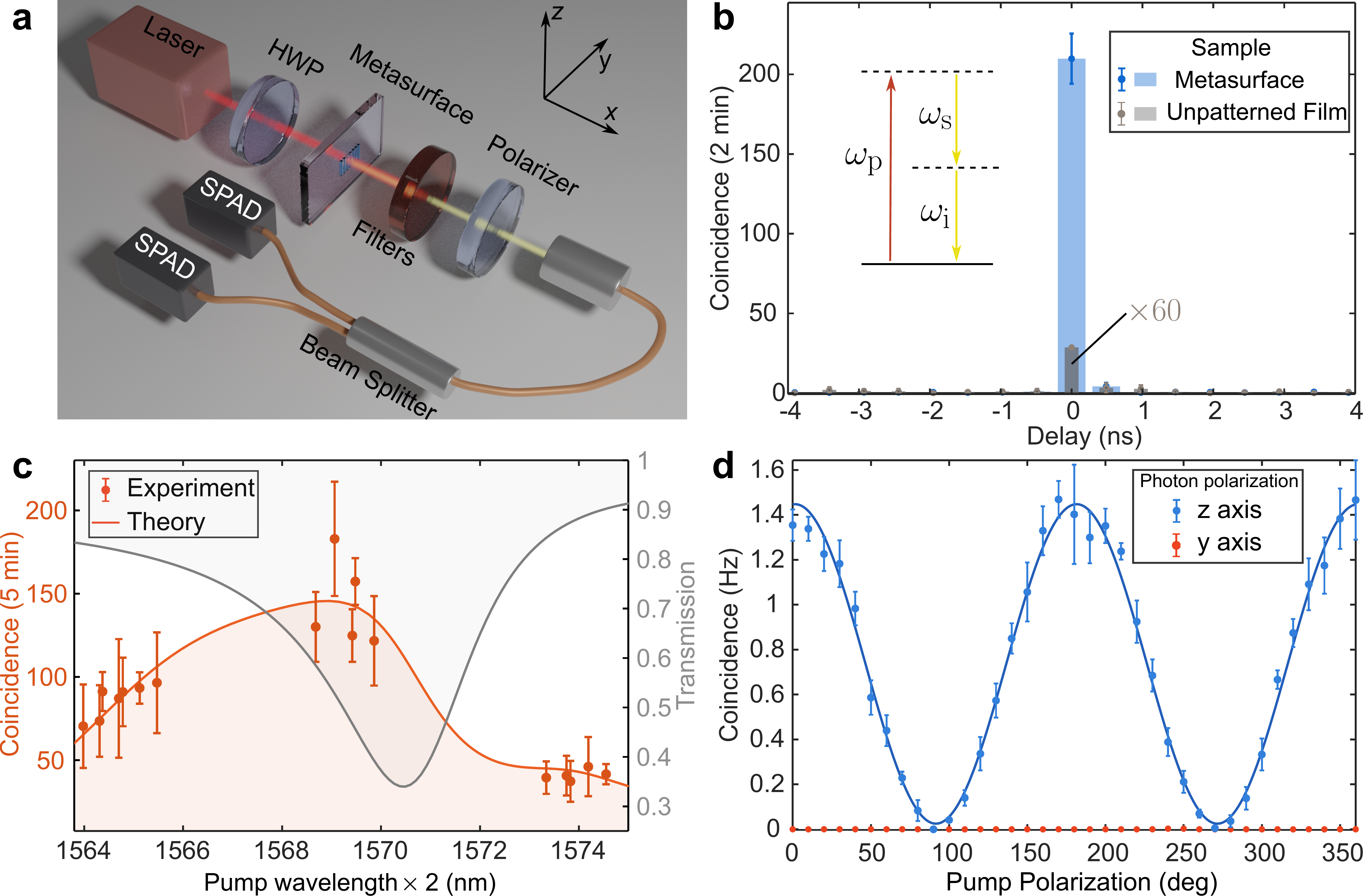} 
\caption{\textbf{Quantum experiment: enhanced generation of photon pairs from the nonlocal metasurface.} 
\textbf{a} Setup for SPDC. A laser beam with a wavelength at 785 nm is focused on a metasurface fabricated on top of a lithium niobate film to produce photon pairs. The photon pairs pass through a 50:50 fiber beam splitter, and their coincidence is then registered by two single-photon detectors. \textbf{b} Coincidence histograms of SPDC from metasurface (blue curve) and unpatterned film (red curve). The red curve is obtained via the integration time of 2 hours which is 60 times longer than the measurement of the blue curve. \textbf{c} Real coincidence as a function of the degenerate signal/idler wavelength which is two times the pump wavelength. The experiment (dots with error bars) shows a good agreement with the theoretical result (solid line). The error bars indicate two standard deviations. The transmission curve (grey) is the same as the one given in Fig.~\ref{fig:classical}\textbf{e}.  \textbf{d} Real coincidence as a function of pump polarization. Points and lines are experimental results and theoretical fitting, respectively, for the photon-pair polarization along the $z$ or $y$ axis of the film as indicated in the inset. The error bars indicate one standard deviation. The pump powers used for panels \textbf{b}, \textbf{c} and \textbf{d} are, 85~mW, 35~mW, 75~mW, respectively.
}
\label{fig:quantum} 
\end{figure*}

We now proceed with the experimental investigations of the quantum photon-pair generation. A setup sketch for characterizing the SPDC is shown in Fig.~\ref{fig:quantum}\textbf{a}. The metasurface is pumped with a continuous-wave laser tunable around the 785~nm wavelength with a beam diameter of $\sim$ \SI{100}{\micro\meter}. The correlations of photon pairs generated from the metasurface are analyzed with a Hanbury Brown-Twiss setup, using a 50:50 multi-mode fiber beam splitter and two single-photon detectors based on InGaAs/InP avalanche photodiodes (see Methods). 

The measured coincidences of photon pairs from the metasurface for different time delays between the detectors are displayed in Fig.~\ref{fig:quantum}\textbf{b}. We obtain a coincidence rate of 1.8~Hz at a pump power of 85~mW, corresponding to a photon-generation efficiency of 21~mHz/mW. The efficiency of single-photon detectors calibrated by the manufacturer is 25\%, and the collection  efficiency for each photon is estimated as 25\%  $\sim$ 30\%, which results in an overall detection efficiency of 0.4\% $\sim$ 0.6\% for the photon pairs. This suggests a photon-pair emission rate of 2.3 $\sim$ 3.5~Hz/mW from the metasurface over a collection angle range calibrated as $\sim\,0.7^\circ$, which closely agrees with the theoretical predictions (see Supplementary Fig.~S3). 

For comparison, we also measure the photon-pair rate from an unpatterned LiNbO$_3$ film under the same experimental conditions, which is found to be 0.047~mHz/mW at zero delay. Based on the peak values of the histograms shown in Fig.~\ref{fig:quantum}\textbf{b}, we identify a rate enhancement of 450 times from the metasurface compared to an unpatterned film, which agrees with the numerical modelling (see Supplementary Fig.~S4\textbf{b}).
Our theory also predicts
that the generated photon pairs have a narrow bandwidth of $\sim 3$~nm and the spectral brightness enhancement at the central wavelength is $\sim$1400~times (see Supplementary Fig.~S4\textbf{a}). 


We confirm that the measured coincidences can be attributed to the generation of non-classical photon pairs by analyzing the second-order correlation function $g^{(2)}(0)$. For the resonantly-enhanced emission from the metasurface, $g^{(2)}(0)$ reaches the value of $\sim$~5000 (see Supplementary Fig.~S8\textbf{a}). The corresponding coincidence to accidental ratio (CAR = $g^{(2)}(0)-1$) is $\sim$~5000, which is over three orders of magnitude larger than the classical bound of 2.
Both the measured rate enhancement and CAR in our metasurface, which are independent of the detector and collection efficiencies, are an order of magnitude higher than previously reported for photon-pair generation from nonlinear metasurfaces with localized Mie resonances~\cite{Santiago-Cruz:2021-4423:NANL}. 
These enhanced rates are
the foundation for the preparation of strongly entangled quantum states with a much higher spectral and spatial brightness~\cite{Parry:2021-55001:ADP, Mazzanti:2022-35006:NJP} compared to localized resonances.

The SPDC efficiency is expected to nontrivially depend on 
the pump wavelength, reflecting the pronounced dispersion of the nonlocal metasurface resonances.
We collect the real coincidence (total peak coincidence with subtracted accidental coincidences) for different pump wavelengths, as shown in Fig.~\ref{fig:quantum}c (red curve), where the horizontal axis indicates the degenerate signal/idler wavelength, i.e. two times of the pump wavelength. 
The data points are taken with different wavelength steps determined by the tuning characteristics of the pump laser
(see Supplementary Sec.~S2.3). The maximum coincidence is found when the degenerate wavelength is slightly blue-detuned from the bright-mode resonance at normal incidence (grey curve), which is 
consistent with the theoretical predictions in Fig.~\ref{fig:theory}\textbf{f}. We fit the experimental data (dots with error bars) to the formulated coupled-mode theory with two free parameters, the collection angle and overall detection efficiency. The fitting results indicate a collection angle of $0.68^\circ$ and a total detection efficiency of 0.4\%, which is a good match with the  experimental parameters presented above.

Experimental measurements confirm that the enhanced photon-pair rate remains close to its maximum over a broad span of pump wavelengths, when the spatial emission is within the range of collection angles.
This efficient spectrum tunability of photon pairs 
is a distinguishing feature of our metasurface with nonlocal resonances, where the transverse phase-matching is satisfied for a continuous set of pump wavelengths, while there are no limitations due to longitudinal phase matching in contrast to bulk nonlinear crystals.


Next, we analyze the dependence of the SPDC on the pump polarization by rotating the half-wave plate (HWP) placed before the metasurface and measuring the photon coincidences with a linear polarizer placed after the metasurface oriented in the $y$ or $z$-direction. We show in Fig.~\ref{fig:quantum}\textbf{d} that the real coincidence count is strongly dependent on the pump polarization when the polarizer is parallel to the $z$-axis (LiNbO$_3$ optical axis); meanwhile, almost no photons are detected at any pump polarization angle when the polarizer is rotated along the $y$ axis. That is, the coincidence rate is maximized when both the pump and emitted photon pairs are polarized along the $z$-axis of the film, when the nonlinear wave mixing is mediated by the strongest quadratic susceptibility tensor component of LiNbO$_3$.
The visibility of the polarization dependence is estimated above 99\%, benefiting from the 
selective resonant enhancement of the SPDC for the photon polarizations along the grating direction according to our metasurface design. 


\begin{figure*}[!b] \centering
\includegraphics[width=1\textwidth]{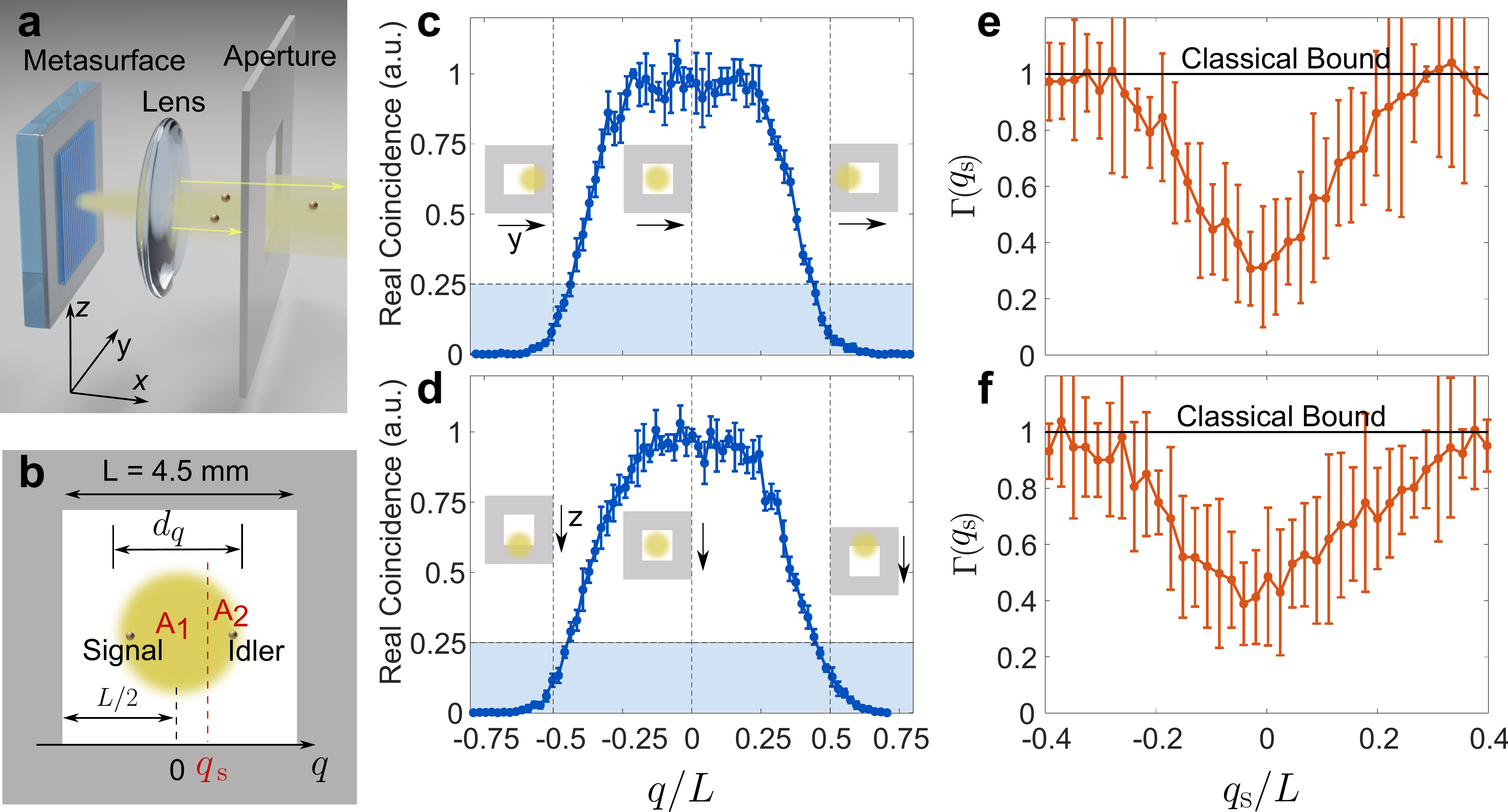} 
\caption{\textbf{Spatial entanglement of photon pairs.} \textbf{a}~A square aperture with a size of $L\times L$ ($L = 4.5$ mm) is introduced after the metasurface to characterize the spatial correlations of photon pairs. \textbf{b}~The aperture size is larger than the profile of the collected beam determined by the size of the fiber collection lens. The position where the centers of the beam and the aperture are aligned is defined as zero. We characterize the correlations for the spatial regions $A_1$ and $A_2$ separated at the position $q_{\rm s}$. \textbf{c,d}~The real coincidence rate of collected photon pairs vs. the aperture position along the $y$ and $z$ directions, normalized to the maximum value. The error bars indicate one standard deviation. \textbf{e,f}~Violation of Cauchy-Schwartz inequality for values below the classical bound. The error bars indicate three standard deviations, corresponding to $>99.7$\% confidence interval.
}
\label{fig:entang} 
\end{figure*}

Finally, we experimentally confirm the spatial entanglement of the photon pairs by selectively blocking photons with an aperture.
Such measurements allow us to reveal the presence of non-classical correlations between the photons at different spatial regions.
%
We use a square aperture placed after the photon emission [Fig.~\ref{fig:entang}\textbf{a}] and translate it transversely in the $y$ or $z$-direction. The width of the aperture is $L=4.5$~mm, which is chosen to be larger than the aperture of the collected photons [Fig.~\ref{fig:entang}\textbf{b}]. The real two-photon coincidences $C(q)$ as a function of the aperture position $q$ ($q = y$ or $q = z$) are presented in Figs.~\ref{fig:entang}\textbf{c}-\textbf{d}. Note that at $q=0$,
the center of the aperture is aligned with the center of the beam. The diameters of the collected beam $d_q$ ($q = y,z$), which are estimated by subtracting the aperture width from the flat plateau region of the correlation curves in Fig.~\ref{fig:entang}\textbf{c}-\textbf{d}, are $d_y=0.55L$ and $d_z=0.65L$ respectively. 

Let us consider a counter-example of classical light correlations, which should satisfy the Cauchy-Schwartz Inequality (CSI) that for our setup can be formulated as (see Supplementary Sec.~S3): 
\begin{equation} \label{eq:CSI}
     \Gamma(q_{s}) \equiv \left(\sqrt{C(q_{s}-L/2)}+ \sqrt{C(q_{s}+L/2)}\right)^2 \geq C(0) \,.
\end{equation}
%
Here, $C(q_{s}-L/2)$ and $C(q_{s}+L/2)$ represent the two-photon coincidences within the two complementary emission regions $A_1$ and $A_2$ separated at the position $q_{\rm s}$, as sketched in Fig.~\ref{fig:entang}\textbf{b}. 
The CSI can be violated for quantum states, when $\Gamma(q_{s}) < C(0)$, indicating {\em non-classical spatial correlations and multi-mode spatial entanglement}~\cite{Wasak:2016-269:QIP}.
The underlying physics 
can be interpreted as the spatial analogue of the photon antibunching effect~\cite{Gatti:2008-251:PROP}. Indeed, the quantities on the left-hand side of Eq.~(\ref{eq:CSI}) are proportional to the self-correlations within the regions $A_1$ or $A_2$, and when their quadratic combination is below the total correlations over the whole beam $C(0)$. This means that the cross-correlations between regions $A_1$ and $A_2$ are higher than what is possible classically.


We identify the violation of CSI by applying Eq.~(\ref{eq:CSI}) to the correlation measurements from Figs.~\ref{fig:entang}\textbf{c}-\textbf{d}, and present the values of $\Gamma(q_{s})$ in Figs.~\ref{fig:entang}\textbf{e}-\textbf{f}. For convenience, we normalize the data such that the plotted value of $C(0)$ is unity, which defines the classical bound.
We see that $\Gamma(q_{s})$ is significantly lower than the classical bound by over three standard deviations for a broad range of spatial region boundaries, $-0.2 < q_s/L < 0.2$. 

Let us estimate the degree of spatial anti-bunching for the case of $q_s=0$, when half of the photons are blocked in either the horizontal or vertical direction (at $q= \pm L/2$ aperture positions). Then, for the photon emission with the spatial reflection symmetry following from the symmetric metasurface design, Eq.~(\ref{eq:CSI}) simplifies to $\Gamma(0) = 4 C(\pm L/2) \geq C(0)$. We see in  Figs.~\ref{fig:entang}\textbf{c}-\textbf{d}
that the coincidence rate drops by more than 4 times (within the area marked by shading) 
at $q= \pm L/2$, providing an additional visual confirmation of CSI violation.
Specifically, the rate reduces by a factor of $\sim$13 along the $y$ ($\sim$6 along $z$) direction at $q = L/2$. 
The corresponding cross-correlations between the half-beam regions, which we estimate using Supplementary Eq.~(S23), are larger than the self-correlations by a factor of $5.5$ for $y$ ($\sim$2 for $z$) direction, confirming the pronounced spatial anti-bunching.

These results demonstrate a noticeable difference between the emission directions. Specifically, the correlation function $\Gamma(q_{s})$ values at $q_{s} = 0$ are $0.3\,C(0)$ and $0.5\,C(0)$ for $y$ and $z$-directions, respectively. The larger CSI violation for the $y$ direction agrees with the stronger anti-bunching estimated above.
This suggests the presence of stronger spatial entanglement along $y$ compared to the $z$ direction, which might be related to the off-normal emission of photons in the $y$ direction according to the theoretical predictions 
shown in Fig.~\ref{fig:theory}\textbf{d}. 

\section*{Discussion}
In summary, we have proposed and experimentally demonstrated that enhanced generation of  quantum photon-pair states can be facilitated through specially designed meta-gratings fabricated on top of a lithium niobate film with a sub-wavelength thickness. The metasurface supports nonlocal resonances that allow transverse phase-matching of spontaneous parametric down-conversion and the simultaneous control of the angular emission pattern over a broad pump wavelength tuning range, while the longitudinal matching requirements are removed due to ultra-small thickness.
These unique features strongly enhance the photon-pair generation rate by over $\sim450$ times compared to unpatterned structures, while the coincidence-to-accidentals ratio reaches $\sim 5000$, demonstrating the high quality of quantum states. 
Our metasurface platform can lead to even higher photon rates and brightness by increasing the quality-factor of optical resonances through improvements in the nanofabrication precision.

The generated photons can be strongly entangled in space, while being almost indistinguishable in other degrees of freedom. In particular, we detect a purely linear polarization state of the photons with a high extinction ratio above 99\%, which is achieved by designing the grating to selectively enhance the electric field component along the optical axis of the nonlinear film. On the other hand, theoretical modelling predicts a near-degenerate narrow emission spectrum of about $\sim 3$~nm. 
%
We experimentally characterized the spatial correlations of photon pairs by partially blocking the emission with an aperture and detected the violation of classical Cauchy-Schwartz inequality, which serves as a criterion of spatial anti-bunching and multi-mode entanglement.
We anticipate that future developments of this ultra-thin platform, including the incorporation of inhomogeneous and two-dimensional meta-grating patterns, can allow even more flexibility in enhancing and shaping the photon emission and entanglement, paving the way towards various applications such as quantum imaging.







\section*{Materials and methods}
{\bf Numerical simulations:}
The simulations of transmission spectra and eigenfrequencies are performed based on the Finite Element Method by using the Comsol Multiphysics software package. The fitting of the CMT parameters and calculation of SPDC rate with CMT are done in Matlab.

{\bf Metasurface fabrication:}
The designed metasurface was fabricated in a cleanroom starting from a LiNbO$_3$ film on quartz substrate from NANOLN. After ultrasonic cleaning, a thin layer of SiO$_2$ with a thickness of ~200 nm was deposited on the sample by plasma-enhanced chemical vapor deposition. Then, PMMA was spin-coated as the resist for the subsequent electron beam lithography. After lithography, a thin layer of nickel with a thickness of 30 nm was coated by electron beam deposition followed by a lift-off process. The nickel pattern was used as the mask for etching of the SiO$_2$ layer by inductively coupled plasma etching. Finally, the residual nickel mask was removed by chemical etching. The size of the grating is $400\mu m\times400\mu m$.

{\bf Second-harmonic characterisation:}
The femtosecond laser (Chameleon Compact OPO, Coherent) used to characterize the SHG from metasurface is set with following parameters: beam diameter of $\sim$ \SI{100}{\micro\meter}; power of \SI{10}{mW}; pulse width of \SI{200}{\femto\second}; linewidth of \SI{23}{nm}; repetition rate of 80 MHz. We employ a lens with a focal length of \SI{150}{mm} to focus the laser and an objective of $\times 20$ to collect the signal. The laser beam transmitting through the metasurface is removed with a short pass filter before the signals are sent to the spectrometer. 

{\bf SPDC experiments:}
The laser exciting the SPDC process is emitted from a Fabry-Perot laser diode (FPL785P, Thorlabs). The laser wavelength can be tuned from 780 nm to 790 nm, with a linewidth of $\sim$ 0.1 nm. A short-pass filter at 850 nm before the metasurface and a long-pass filter at 1100 nm along with a band-pass filter at 1570 nm (with a 50~nm FWHM) after metasurface suppress the fluorescence produced by the metasurface and other optics. Two lenses with a focal length of 100 mm are used to focus the pump beam and collimate the photons emitted from the metasurface. The photon pairs are collected with a multimode fiber and are probabilistically split into two optical paths with a 50:50 fiber beam splitter. The photons are registered with two single-photon detectors based on InGaAs/InP avalanche photodiodes (ID230, IDQ). The detection events are characterized by a time-to-digital converter (ID801, IDQ), which coincidence window is set at 0.486~ns.

\section*{Data availability} 
The data that support the findings of this study are available from the corresponding author upon reasonable request.

\bibliography{mybib.bib,db_art_SPDC_LiNb_1D.bib,mybib_JZ}

\begin{thebibliography}{42}%
\makeatletter
\providecommand \@ifxundefined [1]{%
 \@ifx{#1\undefined}
}%
\providecommand \@ifnum [1]{%
 \ifnum #1\expandafter \@firstoftwo
 \else \expandafter \@secondoftwo
 \fi
}%
\providecommand \@ifx [1]{%
 \ifx #1\expandafter \@firstoftwo
 \else \expandafter \@secondoftwo
 \fi
}%
\providecommand \natexlab [1]{#1}%
\providecommand \enquote  [1]{``#1''}%
\providecommand \bibnamefont  [1]{#1}%
\providecommand \bibfnamefont [1]{#1}%
\providecommand \citenamefont [1]{#1}%
\providecommand \href@noop [0]{\@secondoftwo}%
\providecommand \href [0]{\begingroup \@sanitize@url \@href}%
\providecommand \@href[1]{\@@startlink{#1}\@@href}%
\providecommand \@@href[1]{\endgroup#1\@@endlink}%
\providecommand \@sanitize@url [0]{\catcode `\\12\catcode `\$12\catcode
  `\&12\catcode `\#12\catcode `\^12\catcode `\_12\catcode `\%12\relax}%
\providecommand \@@startlink[1]{}%
\providecommand \@@endlink[0]{}%
\providecommand \url  [0]{\begingroup\@sanitize@url \@url }%
\providecommand \@url [1]{\endgroup\@href {#1}{\urlprefix }}%
\providecommand \urlprefix  [0]{URL }%
\providecommand \Eprint [0]{\href }%
\providecommand \doibase [0]{https://doi.org/}%
\providecommand \selectlanguage [0]{\@gobble}%
\providecommand \bibinfo  [0]{\@secondoftwo}%
\providecommand \bibfield  [0]{\@secondoftwo}%
\providecommand \translation [1]{[#1]}%
\providecommand \BibitemOpen [0]{}%
\providecommand \bibitemStop [0]{}%
\providecommand \bibitemNoStop [0]{.\EOS\space}%
\providecommand \EOS [0]{\spacefactor3000\relax}%
\providecommand \BibitemShut  [1]{\csname bibitem#1\endcsname}%
\let\auto@bib@innerbib\@empty
\bibitem [{\citenamefont {Horodecki}\ \emph {et~al.}(2009)\citenamefont
  {Horodecki}, \citenamefont {Horodecki}, \citenamefont {Horodecki},\ and\
  \citenamefont {Horodecki}}]{Horodecki:2009-865:RMP}%
  \BibitemOpen
  \bibfield  {author} {\bibinfo {author} {\bibfnamefont {R.}~\bibnamefont
  {Horodecki}}, \bibinfo {author} {\bibfnamefont {P.}~\bibnamefont
  {Horodecki}}, \bibinfo {author} {\bibfnamefont {M.}~\bibnamefont
  {Horodecki}},\ and\ \bibinfo {author} {\bibfnamefont {K.}~\bibnamefont
  {Horodecki}},\ }\bibfield  {title} {{\selectlanguage {English}\bibinfo
  {title} {Quantum entanglement}},\ }\href
  {https://doi.org/10.1103/RevModPhys.81.865} {\bibfield  {journal} {\bibinfo
  {journal} {Rev. Mod. Phys.}\ }\textbf {\bibinfo {volume} {81}},\ \bibinfo
  {pages} {865} (\bibinfo {year} {2009})}\BibitemShut {NoStop}%
\bibitem [{\citenamefont {Shih}(2007)}]{Shih:2007-1016:ISQE}%
  \BibitemOpen
  \bibfield  {author} {\bibinfo {author} {\bibfnamefont {Y.~H.}\ \bibnamefont
  {Shih}},\ }\bibfield  {title} {{\selectlanguage {English}\bibinfo {title}
  {Quantum imaging}},\ }\href {https://doi.org/10.1109/JSTQE.2007.902724}
  {\bibfield  {journal} {\bibinfo  {journal} {IEEE J. Sel. Top. Quantum
  Electron.}\ }\textbf {\bibinfo {volume} {13}},\ \bibinfo {pages} {1016}
  (\bibinfo {year} {2007})}\BibitemShut {NoStop}%
\bibitem [{\citenamefont {Gisin}\ and\ \citenamefont
  {Thew}(2007)}]{Gisin:2007-165:NPHOT}%
  \BibitemOpen
  \bibfield  {author} {\bibinfo {author} {\bibfnamefont {N.}~\bibnamefont
  {Gisin}}\ and\ \bibinfo {author} {\bibfnamefont {R.}~\bibnamefont {Thew}},\
  }\bibfield  {title} {{\selectlanguage {English}\bibinfo {title} {Quantum
  communication}},\ }\href {https://doi.org/10.1038/nphoton.2007.22} {\bibfield
   {journal} {\bibinfo  {journal} {Nat. Photon.}\ }\textbf {\bibinfo {volume}
  {1}},\ \bibinfo {pages} {165} (\bibinfo {year} {2007})}\BibitemShut {NoStop}%
\bibitem [{\citenamefont {Bennett}\ and\ \citenamefont
  {DiVincenzo}(2000)}]{Bennett:2000-247:NAT}%
  \BibitemOpen
  \bibfield  {author} {\bibinfo {author} {\bibfnamefont {C.~H.}\ \bibnamefont
  {Bennett}}\ and\ \bibinfo {author} {\bibfnamefont {D.~P.}\ \bibnamefont
  {DiVincenzo}},\ }\bibfield  {title} {{\selectlanguage {English}\bibinfo
  {title} {Quantum information and computation}},\ }\href
  {https://doi.org/10.1038/35005001} {\bibfield  {journal} {\bibinfo  {journal}
  {Nature}\ }\textbf {\bibinfo {volume} {404}},\ \bibinfo {pages} {247}
  (\bibinfo {year} {2000})}\BibitemShut {NoStop}%
\bibitem [{\citenamefont {Klyshko}(1988)}]{Klyshko:1988:PhotonsNonlinear}%
  \BibitemOpen
  \bibfield  {author} {\bibinfo {author} {\bibfnamefont {D.}~\bibnamefont
  {Klyshko}},\ }\href {https://doi.org/10.1201/9780203743508} {\emph {\bibinfo
  {title} {{Photons and Nonlinear Optics}}}}\ (\bibinfo  {publisher} {Gordon
  and Breach},\ \bibinfo {address} {New York},\ \bibinfo {year}
  {1988})\BibitemShut {NoStop}%
\bibitem [{\citenamefont {Chan}\ \emph {et~al.}(2007)\citenamefont {Chan},
  \citenamefont {Torres},\ and\ \citenamefont {Eberly}}]{Chan:2007-50101:PRA}%
  \BibitemOpen
  \bibfield  {author} {\bibinfo {author} {\bibfnamefont {K.~W.}\ \bibnamefont
  {Chan}}, \bibinfo {author} {\bibfnamefont {J.~P.}\ \bibnamefont {Torres}},\
  and\ \bibinfo {author} {\bibfnamefont {J.~H.}\ \bibnamefont {Eberly}},\
  }\bibfield  {title} {{\selectlanguage {English}\bibinfo {title} {Transverse
  entanglement migration in hilbert space}},\ }\href
  {https://doi.org/10.1103/PhysRevA.75.050101} {\bibfield  {journal} {\bibinfo
  {journal} {Phys. Rev. A}\ }\textbf {\bibinfo {volume} {75}},\ \bibinfo
  {pages} {050101} (\bibinfo {year} {2007})}\BibitemShut {NoStop}%
\bibitem [{\citenamefont {Just}\ \emph {et~al.}(2013)\citenamefont {Just},
  \citenamefont {Cavanna}, \citenamefont {Chekhova},\ and\ \citenamefont
  {Leuchs}}]{Just:2013-83015:NJP}%
  \BibitemOpen
  \bibfield  {author} {\bibinfo {author} {\bibfnamefont {F.}~\bibnamefont
  {Just}}, \bibinfo {author} {\bibfnamefont {A.}~\bibnamefont {Cavanna}},
  \bibinfo {author} {\bibfnamefont {M.~V.}\ \bibnamefont {Chekhova}},\ and\
  \bibinfo {author} {\bibfnamefont {G.}~\bibnamefont {Leuchs}},\ }\bibfield
  {title} {{\selectlanguage {English}\bibinfo {title} {Transverse entanglement
  of biphotons}},\ }\href {https://doi.org/10.1088/1367-2630/15/8/083015}
  {\bibfield  {journal} {\bibinfo  {journal} {New J. Phys.}\ }\textbf {\bibinfo
  {volume} {15}},\ \bibinfo {pages} {083015} (\bibinfo {year}
  {2013})}\BibitemShut {NoStop}%
\bibitem [{\citenamefont {Mair}\ \emph {et~al.}(2001)\citenamefont {Mair},
  \citenamefont {Vaziri}, \citenamefont {Weihs},\ and\ \citenamefont
  {Zeilinger}}]{Mair:2001-313:NAT}%
  \BibitemOpen
  \bibfield  {author} {\bibinfo {author} {\bibfnamefont {A.}~\bibnamefont
  {Mair}}, \bibinfo {author} {\bibfnamefont {A.}~\bibnamefont {Vaziri}},
  \bibinfo {author} {\bibfnamefont {G.}~\bibnamefont {Weihs}},\ and\ \bibinfo
  {author} {\bibfnamefont {A.}~\bibnamefont {Zeilinger}},\ }\bibfield  {title}
  {{\selectlanguage {English}\bibinfo {title} {Entanglement of the orbital
  angular momentum states of photons}},\ }\href
  {https://doi.org/10.1038/35085529} {\bibfield  {journal} {\bibinfo  {journal}
  {Nature}\ }\textbf {\bibinfo {volume} {412}},\ \bibinfo {pages} {313}
  (\bibinfo {year} {2001})}\BibitemShut {NoStop}%
\bibitem [{\citenamefont {Law}\ and\ \citenamefont
  {Eberly}(2004)}]{Law:2004-127903:PRL}%
  \BibitemOpen
  \bibfield  {author} {\bibinfo {author} {\bibfnamefont {C.~K.}\ \bibnamefont
  {Law}}\ and\ \bibinfo {author} {\bibfnamefont {J.~H.}\ \bibnamefont
  {Eberly}},\ }\bibfield  {title} {{\selectlanguage {English}\bibinfo {title}
  {Analysis and interpretation of high transverse entanglement in optical
  parametric down conversion}},\ }\href
  {https://doi.org/10.1103/PhysRevLett.92.127903} {\bibfield  {journal}
  {\bibinfo  {journal} {Phys. Rev. Lett.}\ }\textbf {\bibinfo {volume} {92}},\
  \bibinfo {pages} {127903} (\bibinfo {year} {2004})}\BibitemShut {NoStop}%
\bibitem [{\citenamefont {Torres}\ \emph {et~al.}(2003)\citenamefont {Torres},
  \citenamefont {Alexandrescu},\ and\ \citenamefont
  {Torner}}]{Torres:2003-50301:PRA}%
  \BibitemOpen
  \bibfield  {author} {\bibinfo {author} {\bibfnamefont {J.~P.}\ \bibnamefont
  {Torres}}, \bibinfo {author} {\bibfnamefont {A.}~\bibnamefont
  {Alexandrescu}},\ and\ \bibinfo {author} {\bibfnamefont {L.}~\bibnamefont
  {Torner}},\ }\bibfield  {title} {{\selectlanguage {English}\bibinfo {title}
  {Quantum spiral bandwidth of entangled two-photon states}},\ }\href
  {https://doi.org/10.1103/PhysRevA.68.050301} {\bibfield  {journal} {\bibinfo
  {journal} {Phys. Rev. A}\ }\textbf {\bibinfo {volume} {68}},\ \bibinfo
  {pages} {050301} (\bibinfo {year} {2003})}\BibitemShut {NoStop}%
\bibitem [{\citenamefont {Salakhutdinov}\ \emph {et~al.}(2012)\citenamefont
  {Salakhutdinov}, \citenamefont {Eliel},\ and\ \citenamefont
  {Loffler}}]{Salakhutdinov:2012-173604:PRL}%
  \BibitemOpen
  \bibfield  {author} {\bibinfo {author} {\bibfnamefont {V.~D.}\ \bibnamefont
  {Salakhutdinov}}, \bibinfo {author} {\bibfnamefont {E.~R.}\ \bibnamefont
  {Eliel}},\ and\ \bibinfo {author} {\bibfnamefont {W.}~\bibnamefont
  {Loffler}},\ }\bibfield  {title} {{\selectlanguage {English}\bibinfo {title}
  {Full-field quantum correlations of spatially entangled photons}},\ }\href
  {https://doi.org/10.1103/PhysRevLett.108.173604} {\bibfield  {journal}
  {\bibinfo  {journal} {Phys. Rev. Lett.}\ }\textbf {\bibinfo {volume} {108}},\
  \bibinfo {pages} {173604} (\bibinfo {year} {2012})}\BibitemShut {NoStop}%
\bibitem [{\citenamefont {Okoth}\ \emph {et~al.}(2020)\citenamefont {Okoth},
  \citenamefont {Kovlakov}, \citenamefont {Bonsel}, \citenamefont {Cavanna},
  \citenamefont {Straupe}, \citenamefont {Kulik},\ and\ \citenamefont
  {Chekhova}}]{Okoth:2020-11801:PRA}%
  \BibitemOpen
  \bibfield  {author} {\bibinfo {author} {\bibfnamefont {C.}~\bibnamefont
  {Okoth}}, \bibinfo {author} {\bibfnamefont {E.}~\bibnamefont {Kovlakov}},
  \bibinfo {author} {\bibfnamefont {F.}~\bibnamefont {Bonsel}}, \bibinfo
  {author} {\bibfnamefont {A.}~\bibnamefont {Cavanna}}, \bibinfo {author}
  {\bibfnamefont {S.}~\bibnamefont {Straupe}}, \bibinfo {author} {\bibfnamefont
  {S.~P.}\ \bibnamefont {Kulik}},\ and\ \bibinfo {author} {\bibfnamefont
  {M.~V.}\ \bibnamefont {Chekhova}},\ }\bibfield  {title} {{\selectlanguage
  {English}\bibinfo {title} {Idealized {E}instein-{P}odolsky-{R}osen states
  from non-phase-matched parametric down-conversion}},\ }\href
  {https://doi.org/10.1103/PhysRevA.101.011801} {\bibfield  {journal} {\bibinfo
   {journal} {Phys. Rev. A}\ }\textbf {\bibinfo {volume} {101}},\ \bibinfo
  {pages} {011801} (\bibinfo {year} {2020})}\BibitemShut {NoStop}%
\bibitem [{\citenamefont {Huang}\ \emph {et~al.}(2020)\citenamefont {Huang},
  \citenamefont {Zhao}, \citenamefont {Zeng}, \citenamefont {Crunteanu},
  \citenamefont {Shum},\ and\ \citenamefont {Yu}}]{Huang:2020-126101:RPP}%
  \BibitemOpen
  \bibfield  {author} {\bibinfo {author} {\bibfnamefont {T.~Y.}\ \bibnamefont
  {Huang}}, \bibinfo {author} {\bibfnamefont {X.}~\bibnamefont {Zhao}},
  \bibinfo {author} {\bibfnamefont {S.~W.}\ \bibnamefont {Zeng}}, \bibinfo
  {author} {\bibfnamefont {A.}~\bibnamefont {Crunteanu}}, \bibinfo {author}
  {\bibfnamefont {P.~P.}\ \bibnamefont {Shum}},\ and\ \bibinfo {author}
  {\bibfnamefont {N.~F.}\ \bibnamefont {Yu}},\ }\bibfield  {title}
  {{\selectlanguage {English}\bibinfo {title} {Planar nonlinear metasurface
  optics and their applications}},\ }\href
  {https://doi.org/10.1088/1361-6633/abb56e} {\bibfield  {journal} {\bibinfo
  {journal} {Rep. Prog. Phys.}\ }\textbf {\bibinfo {volume} {83}},\ \bibinfo
  {pages} {126101} (\bibinfo {year} {2020})}\BibitemShut {NoStop}%
\bibitem [{\citenamefont {Angelis}\ \emph {et~al.}(2020)\citenamefont
  {Angelis}, \citenamefont {Leo},\ and\ \citenamefont
  {Neshev}}]{DeAngelis:2020:NonlinearMetaOptics}%
  \BibitemOpen
  \bibinfo {editor} {\bibfnamefont {C.~D.}\ \bibnamefont {Angelis}}, \bibinfo
  {editor} {\bibfnamefont {G.}~\bibnamefont {Leo}},\ and\ \bibinfo {editor}
  {\bibfnamefont {D.~N.}\ \bibnamefont {Neshev}},\ eds.,\ \href
  {https://doi.org/10.1201/b22515} {\emph {\bibinfo {title} {Nonlinear
  Meta-Optics}}}\ (\bibinfo  {publisher} {CRC Press},\ \bibinfo {address}
  {London},\ \bibinfo {year} {2020})\BibitemShut {NoStop}%
\bibitem [{\citenamefont {Solntsev}\ \emph {et~al.}(2021)\citenamefont
  {Solntsev}, \citenamefont {Agarwal},\ and\ \citenamefont
  {Kivshar}}]{Solntsev:2021-327:NPHOT}%
  \BibitemOpen
  \bibfield  {author} {\bibinfo {author} {\bibfnamefont {A.~S.}\ \bibnamefont
  {Solntsev}}, \bibinfo {author} {\bibfnamefont {G.~S.}\ \bibnamefont
  {Agarwal}},\ and\ \bibinfo {author} {\bibfnamefont {Y.~Y.}\ \bibnamefont
  {Kivshar}},\ }\bibfield  {title} {{\selectlanguage {English}\bibinfo {title}
  {Metasurfaces for quantum photonics}},\ }\href
  {https://doi.org/10.1038/s41566-021-00793-z} {\bibfield  {journal} {\bibinfo
  {journal} {Nat. Photon.}\ }\textbf {\bibinfo {volume} {15}},\ \bibinfo
  {pages} {327} (\bibinfo {year} {2021})}\BibitemShut {NoStop}%
\bibitem [{\citenamefont {Kuznetsov}\ \emph {et~al.}(2016)\citenamefont
  {Kuznetsov}, \citenamefont {Miroshnichenko}, \citenamefont {Brongersma},
  \citenamefont {Kivshar},\ and\ \citenamefont
  {Luk'yanchuk}}]{Kuznetsov:2016-aag2472:SCI}%
  \BibitemOpen
  \bibfield  {author} {\bibinfo {author} {\bibfnamefont {A.~I.}\ \bibnamefont
  {Kuznetsov}}, \bibinfo {author} {\bibfnamefont {A.~E.}\ \bibnamefont
  {Miroshnichenko}}, \bibinfo {author} {\bibfnamefont {M.~L.}\ \bibnamefont
  {Brongersma}}, \bibinfo {author} {\bibfnamefont {Y.~S.}\ \bibnamefont
  {Kivshar}},\ and\ \bibinfo {author} {\bibfnamefont {B.}~\bibnamefont
  {Luk'yanchuk}},\ }\bibfield  {title} {{\selectlanguage {English}\bibinfo
  {title} {Optically resonant dielectric nanostructures}},\ }\href
  {https://doi.org/10.1126/science.aag2472} {\bibfield  {journal} {\bibinfo
  {journal} {Science}\ }\textbf {\bibinfo {volume} {354}},\ \bibinfo {pages}
  {aag2472} (\bibinfo {year} {2016})}\BibitemShut {NoStop}%
\bibitem [{\citenamefont {Marino}\ \emph {et~al.}(2019)\citenamefont {Marino},
  \citenamefont {Solntsev}, \citenamefont {Xu}, \citenamefont {Gili},
  \citenamefont {Carletti}, \citenamefont {Poddubny}, \citenamefont {Rahmani},
  \citenamefont {Smirnova}, \citenamefont {Chen}, \citenamefont {Lemaitre},
  \citenamefont {Zhang}, \citenamefont {Zayats}, \citenamefont {De~Angelis},
  \citenamefont {Leo}, \citenamefont {Sukhorukov},\ and\ \citenamefont
  {Neshev}}]{Marino:2019-1416:OPT}%
  \BibitemOpen
  \bibfield  {author} {\bibinfo {author} {\bibfnamefont {G.}~\bibnamefont
  {Marino}}, \bibinfo {author} {\bibfnamefont {A.~S.}\ \bibnamefont
  {Solntsev}}, \bibinfo {author} {\bibfnamefont {L.}~\bibnamefont {Xu}},
  \bibinfo {author} {\bibfnamefont {V.~F.}\ \bibnamefont {Gili}}, \bibinfo
  {author} {\bibfnamefont {L.}~\bibnamefont {Carletti}}, \bibinfo {author}
  {\bibfnamefont {A.~N.}\ \bibnamefont {Poddubny}}, \bibinfo {author}
  {\bibfnamefont {M.}~\bibnamefont {Rahmani}}, \bibinfo {author} {\bibfnamefont
  {D.~A.}\ \bibnamefont {Smirnova}}, \bibinfo {author} {\bibfnamefont {H.~T.}\
  \bibnamefont {Chen}}, \bibinfo {author} {\bibfnamefont {A.}~\bibnamefont
  {Lemaitre}}, \bibinfo {author} {\bibfnamefont {G.~Q.}\ \bibnamefont {Zhang}},
  \bibinfo {author} {\bibfnamefont {A.~V.}\ \bibnamefont {Zayats}}, \bibinfo
  {author} {\bibfnamefont {C.}~\bibnamefont {De~Angelis}}, \bibinfo {author}
  {\bibfnamefont {G.}~\bibnamefont {Leo}}, \bibinfo {author} {\bibfnamefont
  {A.~A.}\ \bibnamefont {Sukhorukov}},\ and\ \bibinfo {author} {\bibfnamefont
  {D.~N.}\ \bibnamefont {Neshev}},\ }\bibfield  {title} {{\selectlanguage
  {English}\bibinfo {title} {Spontaneous photon-pair generation from a
  dielectric nanoantenna}},\ }\href {https://doi.org/10.1364/OPTICA.6.001416}
  {\bibfield  {journal} {\bibinfo  {journal} {Optica}\ }\textbf {\bibinfo
  {volume} {6}},\ \bibinfo {pages} {1416} (\bibinfo {year} {2019})}\BibitemShut
  {NoStop}%
\bibitem [{\citenamefont {Santiago-Cruz}\ \emph {et~al.}(2021)\citenamefont
  {Santiago-Cruz}, \citenamefont {Fedotova}, \citenamefont {Sultanov},
  \citenamefont {Weissflog}, \citenamefont {Arslan}, \citenamefont {Younesi},
  \citenamefont {Pertsch}, \citenamefont {Staude}, \citenamefont {Setzpfandt},\
  and\ \citenamefont {Chekhova}}]{Santiago-Cruz:2021-4423:NANL}%
  \BibitemOpen
  \bibfield  {author} {\bibinfo {author} {\bibfnamefont {T.}~\bibnamefont
  {Santiago-Cruz}}, \bibinfo {author} {\bibfnamefont {A.}~\bibnamefont
  {Fedotova}}, \bibinfo {author} {\bibfnamefont {V.}~\bibnamefont {Sultanov}},
  \bibinfo {author} {\bibfnamefont {M.~A.}\ \bibnamefont {Weissflog}}, \bibinfo
  {author} {\bibfnamefont {D.}~\bibnamefont {Arslan}}, \bibinfo {author}
  {\bibfnamefont {M.}~\bibnamefont {Younesi}}, \bibinfo {author} {\bibfnamefont
  {T.}~\bibnamefont {Pertsch}}, \bibinfo {author} {\bibfnamefont
  {I.}~\bibnamefont {Staude}}, \bibinfo {author} {\bibfnamefont
  {F.}~\bibnamefont {Setzpfandt}},\ and\ \bibinfo {author} {\bibfnamefont
  {M.}~\bibnamefont {Chekhova}},\ }\bibfield  {title} {{\selectlanguage
  {English}\bibinfo {title} {Photon pairs from resonant metasurfaces}},\ }\href
  {https://doi.org/10.1021/acs.nanolett.1c01125} {\bibfield  {journal}
  {\bibinfo  {journal} {Nano Lett.}\ }\textbf {\bibinfo {volume} {21}},\
  \bibinfo {pages} {4423} (\bibinfo {year} {2021})}\BibitemShut {NoStop}%
\bibitem [{\citenamefont {Parry}\ \emph {et~al.}(2021)\citenamefont {Parry},
  \citenamefont {Mazzanti}, \citenamefont {Poddubny}, \citenamefont
  {Della~Valle}, \citenamefont {Neshev},\ and\ \citenamefont
  {Sukhorukov}}]{Parry:2021-55001:ADP}%
  \BibitemOpen
  \bibfield  {author} {\bibinfo {author} {\bibfnamefont {M.}~\bibnamefont
  {Parry}}, \bibinfo {author} {\bibfnamefont {A.}~\bibnamefont {Mazzanti}},
  \bibinfo {author} {\bibfnamefont {A.}~\bibnamefont {Poddubny}}, \bibinfo
  {author} {\bibfnamefont {G.}~\bibnamefont {Della~Valle}}, \bibinfo {author}
  {\bibfnamefont {D.~N.}\ \bibnamefont {Neshev}},\ and\ \bibinfo {author}
  {\bibfnamefont {A.~A.}\ \bibnamefont {Sukhorukov}},\ }\bibfield  {title}
  {{\selectlanguage {English}\bibinfo {title} {Enhanced generation of
  nondegenerate photon pairs in nonlinear metasurfaces}},\ }\href
  {https://doi.org/10.1117/1.AP.3.5.055001} {\bibfield  {journal} {\bibinfo
  {journal} {Adv. Photon.}\ }\textbf {\bibinfo {volume} {3}},\ \bibinfo {pages}
  {055001} (\bibinfo {year} {2021})}\BibitemShut {NoStop}%
\bibitem [{\citenamefont {Jin}\ \emph {et~al.}(2021)\citenamefont {Jin},
  \citenamefont {Mishra},\ and\ \citenamefont
  {Argyropoulos}}]{Jin:2021-19903:NASC}%
  \BibitemOpen
  \bibfield  {author} {\bibinfo {author} {\bibfnamefont {B.~Y.}\ \bibnamefont
  {Jin}}, \bibinfo {author} {\bibfnamefont {D.}~\bibnamefont {Mishra}},\ and\
  \bibinfo {author} {\bibfnamefont {C.}~\bibnamefont {Argyropoulos}},\
  }\bibfield  {title} {{\selectlanguage {English}\bibinfo {title} {Efficient
  single-photon pair generation by spontaneous parametric down-conversion in
  nonlinear plasmonic metasurfaces}},\ }\href
  {https://doi.org/10.1039/d1nr05379e} {\bibfield  {journal} {\bibinfo
  {journal} {Nanoscale}\ }\textbf {\bibinfo {volume} {13}},\ \bibinfo {pages}
  {19903} (\bibinfo {year} {2021})}\BibitemShut {NoStop}%
\bibitem [{\citenamefont {Mazzanti}\ \emph {et~al.}(2022)\citenamefont
  {Mazzanti}, \citenamefont {Parry}, \citenamefont {Poddubny}, \citenamefont
  {Della~Valle}, \citenamefont {Neshev},\ and\ \citenamefont
  {Sukhorukov}}]{Mazzanti:2022-35006:NJP}%
  \BibitemOpen
  \bibfield  {author} {\bibinfo {author} {\bibfnamefont {A.}~\bibnamefont
  {Mazzanti}}, \bibinfo {author} {\bibfnamefont {M.}~\bibnamefont {Parry}},
  \bibinfo {author} {\bibfnamefont {A.}~\bibnamefont {Poddubny}}, \bibinfo
  {author} {\bibfnamefont {G.}~\bibnamefont {Della~Valle}}, \bibinfo {author}
  {\bibfnamefont {D.~N.}\ \bibnamefont {Neshev}},\ and\ \bibinfo {author}
  {\bibfnamefont {A.~A.}\ \bibnamefont {Sukhorukov}},\ }\bibfield  {title}
  {{\selectlanguage {English}\bibinfo {title} {Enhanced generation of angle
  correlated photon-pairs in nonlinear metasurfaces}},\ }\href
  {https://doi.org/10.1088/1367-2630/ac599e} {\bibfield  {journal} {\bibinfo
  {journal} {New J. Phys.}\ }\textbf {\bibinfo {volume} {24}},\ \bibinfo
  {pages} {035006} (\bibinfo {year} {2022})}\BibitemShut {NoStop}%
\bibitem [{\citenamefont {Klopfer}\ \emph {et~al.}(2022)\citenamefont
  {Klopfer}, \citenamefont {Dagli}, \citenamefont {III}, \citenamefont
  {Lawrence},\ and\ \citenamefont {Dionne}}]{Klopfer:2020-5127:NANL}%
  \BibitemOpen
  \bibfield  {author} {\bibinfo {author} {\bibfnamefont {E.}~\bibnamefont
  {Klopfer}}, \bibinfo {author} {\bibfnamefont {S.}~\bibnamefont {Dagli}},
  \bibinfo {author} {\bibfnamefont {D.~B.}\ \bibnamefont {III}}, \bibinfo
  {author} {\bibfnamefont {M.}~\bibnamefont {Lawrence}},\ and\ \bibinfo
  {author} {\bibfnamefont {J.~A.}\ \bibnamefont {Dionne}},\ }\bibfield  {title}
  {{\selectlanguage {English}\bibinfo {title} {High-quality-factor
  silicon-on-lithium niobate metasurfaces for electro-optically reconfigurable
  wavefront shaping}},\ }\href
  {https://doi.org/10.1021/10.1021/acs.nanolett.1c04723} {\bibfield  {journal}
  {\bibinfo  {journal} {Nano Lett.}\ }\textbf {\bibinfo {volume} {22}},\
  \bibinfo {pages} {1703} (\bibinfo {year} {2022})}\BibitemShut {NoStop}%
\bibitem [{\citenamefont {Weiss}\ \emph {et~al.}(2022)\citenamefont {Weiss},
  \citenamefont {Frydendahl}, \citenamefont {Bar-David}, \citenamefont
  {Zektzer}, \citenamefont {Edrei}, \citenamefont {Engelberg}, \citenamefont
  {Mazurski}, \citenamefont {Desiatov},\ and\ \citenamefont
  {Levy}}]{Weiss:2022-605:ACSP}%
  \BibitemOpen
  \bibfield  {author} {\bibinfo {author} {\bibfnamefont {A.}~\bibnamefont
  {Weiss}}, \bibinfo {author} {\bibfnamefont {C.}~\bibnamefont {Frydendahl}},
  \bibinfo {author} {\bibfnamefont {J.}~\bibnamefont {Bar-David}}, \bibinfo
  {author} {\bibfnamefont {R.}~\bibnamefont {Zektzer}}, \bibinfo {author}
  {\bibfnamefont {E.}~\bibnamefont {Edrei}}, \bibinfo {author} {\bibfnamefont
  {J.}~\bibnamefont {Engelberg}}, \bibinfo {author} {\bibfnamefont
  {N.}~\bibnamefont {Mazurski}}, \bibinfo {author} {\bibfnamefont
  {B.}~\bibnamefont {Desiatov}},\ and\ \bibinfo {author} {\bibfnamefont
  {U.}~\bibnamefont {Levy}},\ }\bibfield  {title} {{\selectlanguage
  {English}\bibinfo {title} {Tunable metasurface using thin-film lithium
  niobate in the telecom regime}},\ }\href
  {https://doi.org/10.1021/acsphotonics.1c01582} {\bibfield  {journal}
  {\bibinfo  {journal} {ACS Photonics}\ }\textbf {\bibinfo {volume} {9}},\
  \bibinfo {pages} {605} (\bibinfo {year} {2022})}\BibitemShut {NoStop}%
\bibitem [{\citenamefont {Duong}\ \emph {et~al.}(2021)\citenamefont {Duong},
  \citenamefont {Saerens}, \citenamefont {Timpu}, \citenamefont {Buscaglia},
  \citenamefont {Buscaglia}, \citenamefont {Morandi}, \citenamefont {Muller},
  \citenamefont {Maeder}, \citenamefont {Kaufmann}, \citenamefont {Sonltsev},\
  and\ \citenamefont {Grange}}]{Duong:2109.08489:ARXIV}%
  \BibitemOpen
  \bibfield  {author} {\bibinfo {author} {\bibfnamefont {N.~M.~H.}\
  \bibnamefont {Duong}}, \bibinfo {author} {\bibfnamefont {G.}~\bibnamefont
  {Saerens}}, \bibinfo {author} {\bibfnamefont {F.}~\bibnamefont {Timpu}},
  \bibinfo {author} {\bibfnamefont {M.~T.}\ \bibnamefont {Buscaglia}}, \bibinfo
  {author} {\bibfnamefont {V.}~\bibnamefont {Buscaglia}}, \bibinfo {author}
  {\bibfnamefont {A.}~\bibnamefont {Morandi}}, \bibinfo {author} {\bibfnamefont
  {J.~S.}\ \bibnamefont {Muller}}, \bibinfo {author} {\bibfnamefont
  {A.}~\bibnamefont {Maeder}}, \bibinfo {author} {\bibfnamefont
  {F.}~\bibnamefont {Kaufmann}}, \bibinfo {author} {\bibfnamefont
  {A.}~\bibnamefont {Sonltsev}},\ and\ \bibinfo {author} {\bibfnamefont
  {R.}~\bibnamefont {Grange}},\ }\bibfield  {title} {\bibinfo {title}
  {Broadband photon pair generation from a single lithium niobate microcube},\
  }\href {https://doi.org/10.48550/arXiv.2109.08489} {\bibfield  {journal}
  {\bibinfo  {journal} {arXiv}\ }\textbf {\bibinfo {volume} {\mdseries
  2109.08489}} (\bibinfo {year} {2021})}\BibitemShut {NoStop}%
\bibitem [{\citenamefont {Saravi}\ \emph {et~al.}(2021)\citenamefont {Saravi},
  \citenamefont {Pertsch},\ and\ \citenamefont
  {Setzpfandt}}]{Saravi:2021-2100789:ADOM}%
  \BibitemOpen
  \bibfield  {author} {\bibinfo {author} {\bibfnamefont {S.}~\bibnamefont
  {Saravi}}, \bibinfo {author} {\bibfnamefont {T.}~\bibnamefont {Pertsch}},\
  and\ \bibinfo {author} {\bibfnamefont {F.}~\bibnamefont {Setzpfandt}},\
  }\bibfield  {title} {{\selectlanguage {English}\bibinfo {title} {Lithium
  niobate on insulator: An emerging platform for integrated quantum
  photonics}},\ }\href {https://doi.org/10.1002/adom.202100789} {\bibfield
  {journal} {\bibinfo  {journal} {Adv. Opt. Mater.}\ }\textbf {\bibinfo
  {volume} {9}},\ \bibinfo {pages} {2100789} (\bibinfo {year}
  {2021})}\BibitemShut {NoStop}%
\bibitem [{\citenamefont {Zhu}\ \emph {et~al.}(2021{\natexlab{a}})\citenamefont
  {Zhu}, \citenamefont {Shao}, \citenamefont {Yu}, \citenamefont {Cheng},
  \citenamefont {Desiatov}, \citenamefont {Xin}, \citenamefont {Hu},
  \citenamefont {Holzgrafe}, \citenamefont {Ghosh}, \citenamefont
  {Shams-Ansari}, \citenamefont {Puma}, \citenamefont {Sinclair}, \citenamefont
  {Reimer}, \citenamefont {Zhang},\ and\ \citenamefont
  {Loncar}}]{Zhu:2021-242:ADOP}%
  \BibitemOpen
  \bibfield  {author} {\bibinfo {author} {\bibfnamefont {D.}~\bibnamefont
  {Zhu}}, \bibinfo {author} {\bibfnamefont {L.~B.}\ \bibnamefont {Shao}},
  \bibinfo {author} {\bibfnamefont {M.~J.}\ \bibnamefont {Yu}}, \bibinfo
  {author} {\bibfnamefont {R.}~\bibnamefont {Cheng}}, \bibinfo {author}
  {\bibfnamefont {B.}~\bibnamefont {Desiatov}}, \bibinfo {author}
  {\bibfnamefont {C.~J.}\ \bibnamefont {Xin}}, \bibinfo {author} {\bibfnamefont
  {Y.~W.}\ \bibnamefont {Hu}}, \bibinfo {author} {\bibfnamefont
  {J.}~\bibnamefont {Holzgrafe}}, \bibinfo {author} {\bibfnamefont
  {S.}~\bibnamefont {Ghosh}}, \bibinfo {author} {\bibfnamefont
  {A.}~\bibnamefont {Shams-Ansari}}, \bibinfo {author} {\bibfnamefont
  {E.}~\bibnamefont {Puma}}, \bibinfo {author} {\bibfnamefont {N.}~\bibnamefont
  {Sinclair}}, \bibinfo {author} {\bibfnamefont {C.}~\bibnamefont {Reimer}},
  \bibinfo {author} {\bibfnamefont {M.~A.}\ \bibnamefont {Zhang}},\ and\
  \bibinfo {author} {\bibfnamefont {M.}~\bibnamefont {Loncar}},\ }\bibfield
  {title} {{\selectlanguage {English}\bibinfo {title} {Integrated photonics on
  thin-film lithium niobate}},\ }\href {https://doi.org/10.1364/AOP.411024}
  {\bibfield  {journal} {\bibinfo  {journal} {Adv. Opt. Photon.}\ }\textbf
  {\bibinfo {volume} {13}},\ \bibinfo {pages} {242} (\bibinfo {year}
  {2021}{\natexlab{a}})}\BibitemShut {NoStop}%
\bibitem [{\citenamefont {Zhang}\ \emph {et~al.}(2020)\citenamefont {Zhang},
  \citenamefont {Li}, \citenamefont {Liu}, \citenamefont {Qiu}, \citenamefont
  {Sun}, \citenamefont {He},\ and\ \citenamefont {Zhou}}]{Zhang:2020-76:LSA}%
  \BibitemOpen
  \bibfield  {author} {\bibinfo {author} {\bibfnamefont {X.~Y.}\ \bibnamefont
  {Zhang}}, \bibinfo {author} {\bibfnamefont {Q.}~\bibnamefont {Li}}, \bibinfo
  {author} {\bibfnamefont {F.~F.}\ \bibnamefont {Liu}}, \bibinfo {author}
  {\bibfnamefont {M.}~\bibnamefont {Qiu}}, \bibinfo {author} {\bibfnamefont
  {S.~L.}\ \bibnamefont {Sun}}, \bibinfo {author} {\bibfnamefont
  {Q.}~\bibnamefont {He}},\ and\ \bibinfo {author} {\bibfnamefont
  {L.}~\bibnamefont {Zhou}},\ }\bibfield  {title} {{\selectlanguage
  {English}\bibinfo {title} {Controlling angular dispersions in optical
  metasurfaces}},\ }\href {https://doi.org/10.1038/s41377-020-0313-0}
  {\bibfield  {journal} {\bibinfo  {journal} {Light Sci. Appl.}\ }\textbf
  {\bibinfo {volume} {9}},\ \bibinfo {pages} {76} (\bibinfo {year}
  {2020})}\BibitemShut {NoStop}%
\bibitem [{\citenamefont {Song}\ \emph {et~al.}(2021)\citenamefont {Song},
  \citenamefont {van~de Groep}, \citenamefont {Kim},\ and\ \citenamefont
  {Brongersma}}]{Song:2021-1224:NNANO}%
  \BibitemOpen
  \bibfield  {author} {\bibinfo {author} {\bibfnamefont {J.~H.}\ \bibnamefont
  {Song}}, \bibinfo {author} {\bibfnamefont {J.}~\bibnamefont {van~de Groep}},
  \bibinfo {author} {\bibfnamefont {S.~J.}\ \bibnamefont {Kim}},\ and\ \bibinfo
  {author} {\bibfnamefont {M.~L.}\ \bibnamefont {Brongersma}},\ }\bibfield
  {title} {{\selectlanguage {English}\bibinfo {title} {Non-local metasurfaces
  for spectrally decoupled wavefront manipulation and eye tracking}},\ }\href
  {https://doi.org/10.1038/s41565-021-00967-4} {\bibfield  {journal} {\bibinfo
  {journal} {Nat. Nanotech.}\ }\textbf {\bibinfo {volume} {16}},\ \bibinfo
  {pages} {1224} (\bibinfo {year} {2021})}\BibitemShut {NoStop}%
\bibitem [{\citenamefont {Kwon}\ \emph {et~al.}(2018)\citenamefont {Kwon},
  \citenamefont {Sounas}, \citenamefont {Cordaro}, \citenamefont {Polman},\
  and\ \citenamefont {Alu}}]{Kwon:2018-173004:PRL}%
  \BibitemOpen
  \bibfield  {author} {\bibinfo {author} {\bibfnamefont {H.}~\bibnamefont
  {Kwon}}, \bibinfo {author} {\bibfnamefont {D.}~\bibnamefont {Sounas}},
  \bibinfo {author} {\bibfnamefont {A.}~\bibnamefont {Cordaro}}, \bibinfo
  {author} {\bibfnamefont {A.}~\bibnamefont {Polman}},\ and\ \bibinfo {author}
  {\bibfnamefont {A.}~\bibnamefont {Alu}},\ }\bibfield  {title}
  {{\selectlanguage {English}\bibinfo {title} {Nonlocal metasurfaces for
  optical signal processing}},\ }\href
  {https://doi.org/10.1103/PhysRevLett.121.173004} {\bibfield  {journal}
  {\bibinfo  {journal} {Phys. Rev. Lett.}\ }\textbf {\bibinfo {volume} {121}},\
  \bibinfo {pages} {173004} (\bibinfo {year} {2018})}\BibitemShut {NoStop}%
\bibitem [{\citenamefont {Wang}\ and\ \citenamefont
  {Magnusson}(1993)}]{Wang:1993-2606:AOP}%
  \BibitemOpen
  \bibfield  {author} {\bibinfo {author} {\bibfnamefont {S.~S.}\ \bibnamefont
  {Wang}}\ and\ \bibinfo {author} {\bibfnamefont {R.}~\bibnamefont
  {Magnusson}},\ }\bibfield  {title} {{\selectlanguage {English}\bibinfo
  {title} {Theory and applications of guided-mode resonance filters}},\ }\href
  {https://doi.org/10.1364/AO.32.002606} {\bibfield  {journal} {\bibinfo
  {journal} {Appl. Optics}\ }\textbf {\bibinfo {volume} {32}},\ \bibinfo
  {pages} {2606} (\bibinfo {year} {1993})}\BibitemShut {NoStop}%
\bibitem [{\citenamefont {Quaranta}\ \emph {et~al.}(2018)\citenamefont
  {Quaranta}, \citenamefont {Basset}, \citenamefont {Martin},\ and\
  \citenamefont {Gallinet}}]{Quaranta:2018-1800017:LPR}%
  \BibitemOpen
  \bibfield  {author} {\bibinfo {author} {\bibfnamefont {G.}~\bibnamefont
  {Quaranta}}, \bibinfo {author} {\bibfnamefont {G.}~\bibnamefont {Basset}},
  \bibinfo {author} {\bibfnamefont {O.~J.~F.}\ \bibnamefont {Martin}},\ and\
  \bibinfo {author} {\bibfnamefont {B.}~\bibnamefont {Gallinet}},\ }\bibfield
  {title} {{\selectlanguage {English}\bibinfo {title} {Recent advances in
  resonant waveguide gratings}},\ }\href
  {https://doi.org/10.1002/lpor.201800017} {\bibfield  {journal} {\bibinfo
  {journal} {Laser Photon. Rev.}\ }\textbf {\bibinfo {volume} {12}},\ \bibinfo
  {pages} {1800017} (\bibinfo {year} {2018})}\BibitemShut {NoStop}%
\bibitem [{\citenamefont {Hsu}\ \emph {et~al.}(2016)\citenamefont {Hsu},
  \citenamefont {Zhen}, \citenamefont {Stone}, \citenamefont {Joannopoulos},\
  and\ \citenamefont {Soljacic}}]{Hsu:2016-16048:NRM}%
  \BibitemOpen
  \bibfield  {author} {\bibinfo {author} {\bibfnamefont {C.~W.}\ \bibnamefont
  {Hsu}}, \bibinfo {author} {\bibfnamefont {B.}~\bibnamefont {Zhen}}, \bibinfo
  {author} {\bibfnamefont {A.~D.}\ \bibnamefont {Stone}}, \bibinfo {author}
  {\bibfnamefont {J.~D.}\ \bibnamefont {Joannopoulos}},\ and\ \bibinfo {author}
  {\bibfnamefont {M.}~\bibnamefont {Soljacic}},\ }\bibfield  {title}
  {{\selectlanguage {English}\bibinfo {title} {Bound states in the
  continuum}},\ }\href {https://doi.org/10.1038/natrevmats.2016.48} {\bibfield
  {journal} {\bibinfo  {journal} {Nat. Rev. Mater.}\ }\textbf {\bibinfo
  {volume} {1}},\ \bibinfo {pages} {16048} (\bibinfo {year}
  {2016})}\BibitemShut {NoStop}%
\bibitem [{\citenamefont {Suh}\ \emph {et~al.}(2004)\citenamefont {Suh},
  \citenamefont {Wang},\ and\ \citenamefont {Fan}}]{Suh:2004-1511:IQE}%
  \BibitemOpen
  \bibfield  {author} {\bibinfo {author} {\bibfnamefont {W.}~\bibnamefont
  {Suh}}, \bibinfo {author} {\bibfnamefont {Z.}~\bibnamefont {Wang}},\ and\
  \bibinfo {author} {\bibfnamefont {S.~H.}\ \bibnamefont {Fan}},\ }\bibfield
  {title} {{\selectlanguage {English}\bibinfo {title} {Temporal coupled-mode
  theory and the presence of non-orthogonal modes in lossless multimode
  cavities}},\ }\href {https://doi.org/10.1109/JQE.2004.834773} {\bibfield
  {journal} {\bibinfo  {journal} {IEEE J. Quantum Electron.}\ }\textbf
  {\bibinfo {volume} {40}},\ \bibinfo {pages} {1511} (\bibinfo {year}
  {2004})}\BibitemShut {NoStop}%
\bibitem [{\citenamefont {Sun}\ \emph {et~al.}(2022)\citenamefont {Sun},
  \citenamefont {Jiang}, \citenamefont {Bykov}, \citenamefont {Van},
  \citenamefont {Levy}, \citenamefont {Cai},\ and\ \citenamefont
  {Han}}]{Sun:2201.08156:ARXIV}%
  \BibitemOpen
  \bibfield  {author} {\bibinfo {author} {\bibfnamefont {K.}~\bibnamefont
  {Sun}}, \bibinfo {author} {\bibfnamefont {H.}~\bibnamefont {Jiang}}, \bibinfo
  {author} {\bibfnamefont {D.~A.}\ \bibnamefont {Bykov}}, \bibinfo {author}
  {\bibfnamefont {V.}~\bibnamefont {Van}}, \bibinfo {author} {\bibfnamefont
  {U.}~\bibnamefont {Levy}}, \bibinfo {author} {\bibfnamefont {Y.}~\bibnamefont
  {Cai}},\ and\ \bibinfo {author} {\bibfnamefont {Z.}~\bibnamefont {Han}},\
  }\bibfield  {title} {\bibinfo {title} {One-dimensional bound states in the
  continuum in the w-k space for nonlinear optical applications},\ }\href
  {https://doi.org/10.48550/arXiv.2201.08156} {\bibfield  {journal} {\bibinfo
  {journal} {arXiv}\ }\textbf {\bibinfo {volume} {\mdseries 2201.08156}}
  (\bibinfo {year} {2022})}\BibitemShut {NoStop}%
\bibitem [{\citenamefont {Ma}\ \emph {et~al.}(2020)\citenamefont {Ma},
  \citenamefont {Chen}, \citenamefont {Ren}, \citenamefont {Wu}, \citenamefont
  {Cai},\ and\ \citenamefont {Xu}}]{Ma:2020-145:OL}%
  \BibitemOpen
  \bibfield  {author} {\bibinfo {author} {\bibfnamefont {J.~J.}\ \bibnamefont
  {Ma}}, \bibinfo {author} {\bibfnamefont {J.~X.}\ \bibnamefont {Chen}},
  \bibinfo {author} {\bibfnamefont {M.~X.}\ \bibnamefont {Ren}}, \bibinfo
  {author} {\bibfnamefont {W.}~\bibnamefont {Wu}}, \bibinfo {author}
  {\bibfnamefont {W.}~\bibnamefont {Cai}},\ and\ \bibinfo {author}
  {\bibfnamefont {J.~J.}\ \bibnamefont {Xu}},\ }\bibfield  {title}
  {{\selectlanguage {English}\bibinfo {title} {Second-harmonic generation and
  its nonlinear depolarization from lithium niobate thin films}},\ }\href
  {https://doi.org/10.1364/OL.45.000145} {\bibfield  {journal} {\bibinfo
  {journal} {Opt. Lett.}\ }\textbf {\bibinfo {volume} {45}},\ \bibinfo {pages}
  {145} (\bibinfo {year} {2020})}\BibitemShut {NoStop}%
\bibitem [{\citenamefont {Zhu}\ \emph {et~al.}(2021{\natexlab{b}})\citenamefont
  {Zhu}, \citenamefont {Zhang}, \citenamefont {Xie}, \citenamefont {Chen},
  \citenamefont {Gong}, \citenamefont {Wu}, \citenamefont {Cai}, \citenamefont
  {Zhang}, \citenamefont {Ren},\ and\ \citenamefont {Xu}}]{ZhuInfrared2021}%
  \BibitemOpen
  \bibfield  {author} {\bibinfo {author} {\bibfnamefont {Z.}~\bibnamefont
  {Zhu}}, \bibinfo {author} {\bibfnamefont {D.}~\bibnamefont {Zhang}}, \bibinfo
  {author} {\bibfnamefont {F.}~\bibnamefont {Xie}}, \bibinfo {author}
  {\bibfnamefont {J.}~\bibnamefont {Chen}}, \bibinfo {author} {\bibfnamefont
  {S.}~\bibnamefont {Gong}}, \bibinfo {author} {\bibfnamefont {W.}~\bibnamefont
  {Wu}}, \bibinfo {author} {\bibfnamefont {W.}~\bibnamefont {Cai}}, \bibinfo
  {author} {\bibfnamefont {X.}~\bibnamefont {Zhang}}, \bibinfo {author}
  {\bibfnamefont {M.}~\bibnamefont {Ren}},\ and\ \bibinfo {author}
  {\bibfnamefont {J.}~\bibnamefont {Xu}},\ }\bibfield  {title} {\bibinfo
  {title} {Infrared full-{{Stokes}} polarimetry by parametric up-conversion},\
  }\href@noop {} {\bibfield  {journal} {\bibinfo  {journal} {arXiv:2110.10506
  [physics]}\ } (\bibinfo {year} {2021}{\natexlab{b}})},\ \Eprint
  {https://arxiv.org/abs/2110.10506} {arXiv:2110.10506 [physics]} \BibitemShut
  {NoStop}%
\bibitem [{\citenamefont {Kim}\ and\ \citenamefont
  {Rim}(2018)}]{Kim:2018-4769:ACSP}%
  \BibitemOpen
  \bibfield  {author} {\bibinfo {author} {\bibfnamefont {K.~H.}\ \bibnamefont
  {Kim}}\ and\ \bibinfo {author} {\bibfnamefont {W.~S.}\ \bibnamefont {Rim}},\
  }\bibfield  {title} {{\selectlanguage {English}\bibinfo {title} {Anapole
  resonances facilitated by high-index contrast between substrate and
  dielectric nanodisk enhance vacuum ultraviolet generation}},\ }\href
  {https://doi.org/10.1021/acsphotonics.8b01287} {\bibfield  {journal}
  {\bibinfo  {journal} {ACS Photonics}\ }\textbf {\bibinfo {volume} {5}},\
  \bibinfo {pages} {4769} (\bibinfo {year} {2018})}\BibitemShut {NoStop}%
\bibitem [{\citenamefont {Fedotova}\ \emph {et~al.}(2020)\citenamefont
  {Fedotova}, \citenamefont {Younesi}, \citenamefont {Sautter}, \citenamefont
  {Vaskin}, \citenamefont {Lochner}, \citenamefont {Steinert}, \citenamefont
  {Geiss}, \citenamefont {Pertsch}, \citenamefont {Staude},\ and\ \citenamefont
  {Setzpfandt}}]{Fedotova:2020-8608:NANL}%
  \BibitemOpen
  \bibfield  {author} {\bibinfo {author} {\bibfnamefont {A.}~\bibnamefont
  {Fedotova}}, \bibinfo {author} {\bibfnamefont {M.}~\bibnamefont {Younesi}},
  \bibinfo {author} {\bibfnamefont {J.}~\bibnamefont {Sautter}}, \bibinfo
  {author} {\bibfnamefont {A.}~\bibnamefont {Vaskin}}, \bibinfo {author}
  {\bibfnamefont {F.~J.~F.}\ \bibnamefont {Lochner}}, \bibinfo {author}
  {\bibfnamefont {M.}~\bibnamefont {Steinert}}, \bibinfo {author}
  {\bibfnamefont {R.}~\bibnamefont {Geiss}}, \bibinfo {author} {\bibfnamefont
  {T.}~\bibnamefont {Pertsch}}, \bibinfo {author} {\bibfnamefont
  {I.}~\bibnamefont {Staude}},\ and\ \bibinfo {author} {\bibfnamefont
  {F.}~\bibnamefont {Setzpfandt}},\ }\bibfield  {title} {{\selectlanguage
  {English}\bibinfo {title} {Second-harmonic generation in resonant nonlinear
  metasurfaces based on lithium niobate}},\ }\href
  {https://doi.org/10.1021/acs.nanolett.0c03290} {\bibfield  {journal}
  {\bibinfo  {journal} {Nano Lett.}\ }\textbf {\bibinfo {volume} {20}},\
  \bibinfo {pages} {8608} (\bibinfo {year} {2020})}\BibitemShut {NoStop}%
\bibitem [{\citenamefont {Carletti}\ \emph {et~al.}(2021)\citenamefont
  {Carletti}, \citenamefont {Zilli}, \citenamefont {Moia}, \citenamefont
  {Toma}, \citenamefont {Finazzi}, \citenamefont {De~Angelis}, \citenamefont
  {Neshev},\ and\ \citenamefont {Celebrano}}]{Carletti:2021-731:ACSP}%
  \BibitemOpen
  \bibfield  {author} {\bibinfo {author} {\bibfnamefont {L.}~\bibnamefont
  {Carletti}}, \bibinfo {author} {\bibfnamefont {A.}~\bibnamefont {Zilli}},
  \bibinfo {author} {\bibfnamefont {F.}~\bibnamefont {Moia}}, \bibinfo {author}
  {\bibfnamefont {A.}~\bibnamefont {Toma}}, \bibinfo {author} {\bibfnamefont
  {M.}~\bibnamefont {Finazzi}}, \bibinfo {author} {\bibfnamefont
  {C.}~\bibnamefont {De~Angelis}}, \bibinfo {author} {\bibfnamefont {D.~N.}\
  \bibnamefont {Neshev}},\ and\ \bibinfo {author} {\bibfnamefont
  {M.}~\bibnamefont {Celebrano}},\ }\bibfield  {title} {{\selectlanguage
  {English}\bibinfo {title} {Steering and encoding the polarization of the
  second harmonic in the visible with a monolithic {{LiNbO}$_3$}
  metasurface}},\ }\href {https://doi.org/10.1021/acsphotonics.1c00026}
  {\bibfield  {journal} {\bibinfo  {journal} {ACS Photonics}\ }\textbf
  {\bibinfo {volume} {8}},\ \bibinfo {pages} {731} (\bibinfo {year}
  {2021})}\BibitemShut {NoStop}%
\bibitem [{\citenamefont {Ma}\ \emph {et~al.}(2021)\citenamefont {Ma},
  \citenamefont {Xie}, \citenamefont {Chen}, \citenamefont {Chen},
  \citenamefont {Wu}, \citenamefont {Liu}, \citenamefont {Chen}, \citenamefont
  {Cai}, \citenamefont {Ren},\ and\ \citenamefont {Xu}}]{Ma:2021-2000521:LPR}%
  \BibitemOpen
  \bibfield  {author} {\bibinfo {author} {\bibfnamefont {J.~J.}\ \bibnamefont
  {Ma}}, \bibinfo {author} {\bibfnamefont {F.}~\bibnamefont {Xie}}, \bibinfo
  {author} {\bibfnamefont {W.~J.}\ \bibnamefont {Chen}}, \bibinfo {author}
  {\bibfnamefont {J.~X.}\ \bibnamefont {Chen}}, \bibinfo {author}
  {\bibfnamefont {W.}~\bibnamefont {Wu}}, \bibinfo {author} {\bibfnamefont
  {W.}~\bibnamefont {Liu}}, \bibinfo {author} {\bibfnamefont {Y.~T.}\
  \bibnamefont {Chen}}, \bibinfo {author} {\bibfnamefont {W.}~\bibnamefont
  {Cai}}, \bibinfo {author} {\bibfnamefont {M.~X.}\ \bibnamefont {Ren}},\ and\
  \bibinfo {author} {\bibfnamefont {J.~J.}\ \bibnamefont {Xu}},\ }\bibfield
  {title} {{\selectlanguage {English}\bibinfo {title} {Nonlinear lithium
  niobate metasurfaces for second harmonic generation}},\ }\href
  {https://doi.org/10.1002/lpor.202000521} {\bibfield  {journal} {\bibinfo
  {journal} {Laser Photon. Rev.}\ }\textbf {\bibinfo {volume} {15}},\ \bibinfo
  {pages} {2000521} (\bibinfo {year} {2021})}\BibitemShut {NoStop}%
\bibitem [{\citenamefont {Wasak}\ \emph {et~al.}(2016)\citenamefont {Wasak},
  \citenamefont {Szankowski}, \citenamefont {Trippenbach},\ and\ \citenamefont
  {Chwedenczuk}}]{Wasak:2016-269:QIP}%
  \BibitemOpen
  \bibfield  {author} {\bibinfo {author} {\bibfnamefont {T.}~\bibnamefont
  {Wasak}}, \bibinfo {author} {\bibfnamefont {P.}~\bibnamefont {Szankowski}},
  \bibinfo {author} {\bibfnamefont {M.}~\bibnamefont {Trippenbach}},\ and\
  \bibinfo {author} {\bibfnamefont {J.}~\bibnamefont {Chwedenczuk}},\
  }\bibfield  {title} {{\selectlanguage {English}\bibinfo {title}
  {{C}auchy-{S}chwarz inequality for general measurements as an entanglement
  criterion}},\ }\href {https://doi.org/10.1007/s11128-015-1181-z} {\bibfield
  {journal} {\bibinfo  {journal} {Quantum Inf. Process.}\ }\textbf {\bibinfo
  {volume} {15}},\ \bibinfo {pages} {269} (\bibinfo {year} {2016})}\BibitemShut
  {NoStop}%
\bibitem [{\citenamefont {Gatti}\ \emph {et~al.}(2008)\citenamefont {Gatti},
  \citenamefont {Brambilla},\ and\ \citenamefont
  {Lugiato}}]{Gatti:2008-251:PROP}%
  \BibitemOpen
  \bibfield  {author} {\bibinfo {author} {\bibfnamefont {A.}~\bibnamefont
  {Gatti}}, \bibinfo {author} {\bibfnamefont {E.}~\bibnamefont {Brambilla}},\
  and\ \bibinfo {author} {\bibfnamefont {L.}~\bibnamefont {Lugiato}},\
  }\bibfield  {title} {{\selectlanguage {English}\bibinfo {title} {Quantum
  imaging}},\ }\href {https://doi.org/10.1016/S0079-6638(07)51005-X} {\bibfield
   {journal} {\bibinfo  {journal} {Prog. Optics}\ }\textbf {\bibinfo {volume}
  {51}},\ \bibinfo {pages} {251} (\bibinfo {year} {2008})}\BibitemShut
  {NoStop}%
\end{thebibliography}%


%
%

\section*{Acknowledgments}
We acknowledge the support by the Australian Research Council (DP190101559, CE200100010).

\section*{Author contributions}
A.A.S., L.X., and D.N.N. developed the theoretical concept. J.Z. and M.P. performed the numerical modelling. J.Z. fabricated the sample and performed the linear transmission measurement. J.Y., M.C., and R.C. developed the experimental setup, performed the measurements and processed the data. A.A.S. and D.N.N. supervised the project. All authors analysed the results and co-wrote the paper.

\section*{Competing interests}
The authors declare no competing interests.

\section*{Additional information}
Supplementary information is available for this paper.

\end{document}